\documentclass[11pt, a4paper]{article}
\pdfoutput=1
\usepackage{jheppub}

\usepackage{latexsym, graphicx}
\usepackage{amsmath, amssymb}
\usepackage{color}

\title{Unruh effect under non-equilibrium conditions: \\
Oscillatory motion of an Unruh-DeWitt detector}

\author[a]{Jason Doukas}
\author[b,1]{Shih-Yuin Lin}
\author[c]{B. L. Hu}
\author[d]{and Robert B. Mann}

\affiliation[a]{School of Mathematical Sciences, University of Nottingham, Nottingham NG7 2RD,
United Kingdom}

\affiliation[b]{Department of Physics, National Changhua University of Education, Changhua 50007, Taiwan
\note{Corresponding author.}}

\affiliation[c]{Maryland Center for Fundamental Physics and Joint Quantum Institute,
University of Maryland, College Park, Maryland 20742-4111, USA} 

\affiliation[d]{Department of Physics and Astronomy, University of Waterloo,\\
Waterloo, Canada N2L 3G1}

\emailAdd{jason.doukas@nottingham.ac.uk}
\emailAdd{sylin@cc.ncue.edu.tw}
\emailAdd{blhu@umd.edu}
\emailAdd{rbmann@sciborg.uwaterloo.ca}

\abstract{The Unruh effect refers to the thermal fluctuations a detector experiences while undergoing linear motion with uniform acceleration in a Minkowski vacuum. This thermality can be demonstrated by tracing the vacuum state of the field over the modes beyond the accelerated detector's event horizon. However, the event horizon  is well-defined only if the detector moves with eternal uniform linear acceleration. This idealized condition cannot be fulfilled in realistic situations when the motion unavoidably involves periods of non-uniform acceleration. Many experimental proposals to test the Unruh effect are of this nature. Often circular or oscillatory motion, which lacks an obvious geometric description, is considered in such proposals. The proper perspective for theoretically going beyond, or experimentally testing, the Unruh-Hawking effect in these more general conditions has to be offered by concepts and techniques in non-equilibrium quantum field theory. In this paper we provide a detailed analysis of how an Unruh-DeWitt detector undergoing oscillatory motion responds to the fluctuations of a quantum field. Numerical results for the late-time temperatures of the oscillating detector are presented. We comment on the digressions of these results from what one would obtain from a naive application of Unruh's result.} 

\keywords{quantum dissipative system, boundary quantum field theory, black holes.}

\arxivnumber{1307.4360}

\begin{document}

\maketitle

\section{Introduction}

The Unruh effect \cite{Unr76,DeW79} attests that a detector 
(an atom or a particle with internal degrees of freedom coupled to the field to be detected) 
moving with a uniform proper acceleration $a$ sees the vacuum state of the quantum field as a thermal bath with the Unruh temperature:
\begin{equation}\label{Unruhequation}
T_{U}=\hbar a/2\pi k_{B}
\end{equation}
(setting the speed of light $c= 1$, as we do throughout this paper). Unruh's classic paper offers an illustrative analog for the Hawking effect in black holes \cite{Haw75}, often explained in terms of the geometric notion of an event horizon.  However, there is no horizon for detectors undergoing non-uniform or finite-time accelerations. One is naturally led to wonder about the existence and robustness of the Unruh effect in situations outside of its original ideal setting. Fortunately, the Unruh effect may be understood purely as a kinematic effect, which was explicitly demonstrated \cite{RHA,RHK,JH1,JHFoP} using the influence functional (Feynman-Vernon) or closed-time-path (Schwinger-Keldysh) formalisms \cite{ifctp,cgea,CalHu08}. In non-uniform motion the detector senses the field via a non-thermal spectral response, with the degree of non-thermality governed by the deviation from uniform acceleration \cite{RHK}.

In fact, a detector that is not uniformly accelerated at all times, such as one in circular or oscillatory motion, 
is in a non-equilibrium state. This was emphasized in \cite{HJ00}, where a theoretical framework for treating the stochastic dynamics of particles interacting with quantum fields under non-equilibrium conditions was developed. Under general conditions, according to Hu and Johnson \cite{HJ00} the moving particle/detector will register a colored noise, which turns thermal only in limiting conditions such as linear uniform acceleration where the Unruh effect shows.

Details of these non-thermal deviations for non-uniformly accelerated detectors are very relevant to experimental proposals of the Unruh effect \cite{Capri00,OtherExpProp,CT99,MFM11}. For obvious practical reasons the proposed experiments are often framed in the setting of circular or oscillatory motion. Notably, the observed spin depolarization of electrons undergoing circular motion was given an Unruh effect interpretation by Bell and Leinaas \cite{BL83, BL87, Len00, Unr98} and Chen and Tajima \cite{CT99} have proposed measuring the Unruh temperature of the radiation from oscillating electrons in ultra strong lasers. See also \cite{Doukas} in which the resource of spin-entanglement was investigated in accelerated electron systems.  

A number of authors have previously considered the response functions 
for the Unruh-DeWitt detectors moving with finite-time \cite{SS92, SP96, Sc04, LS06, Sa07}
and time-varying accelerations \cite{OM07, KP10, BV12} in first-order time-dependent perturbation theory (TDPT), 
though the TDPT will only be of limited validity when the detector motion is non-stationary.
In our view the time is ripe to go beyond the TDPT to analyze problems like that of the oscillating detector, 
where the equilibrium condition does not exist. To do so, 
we consider an atom-like detector, modelled by a harmonic oscillator, that is linearly coupled to the field.
Such a system admits formally exact solutions to the evolution of the two-point correlators of the system and field \cite{LH06,LH07}. Focussing on Gaussian states of the detector and field, the complete dynamics of the system and field can be obtained.  The formal solutions to the two-point correlators  involve double integrals which for any type of detector motion can be calculated numerically. For these systems in Gaussian states the full information between the detector and the field can be obtained and the non-equilibrium effects analyzed. 

In this paper we will explore the extent to which the Unruh temperature could be interpreted as a temperature in non-equilibrium conditions by looking at three types of relativistic linear oscillatory motion: 1) the worldlines of Sinusoidal Motion (SM) formed by the restriction of the circular worldline in the plane to only one of its directions, 2) the worldlines of the electrons in intense lasers in the experiment proposed by Chen and Tajima (CT)  \cite{CT99}, in which the {\it directional} proper acceleration 
\footnote{We found it is convenient to define ``directional proper acceleration"
as the proper acceleration (positive definite) multiplied by the sign of $a^3$ (the $z^3$-component of the 4-acceleration)
for the detector in one-dimensional oscillatory motion.}
of the detector is sinusoidal, and 3) the worldlines consisting of periods of Alternating (in direction) Uniform Acceleration (AUA). Although of significant practical importance, oscillating worldlines have not significantly been investigated previously in the literature. One of the reasons for this is that the noise seen by a detector moving under such motion is time-dependent (non-stationary), making the computations hard and results difficult to interpret. In our work here, in addition to focussing on the kinematical effects on the vacuum fluctuations we also consider two additional aspects that impart  non-equilibrium characteristics to the system: one pertains to the coupling strength, the other to the switch-on time. Strong coupling between the oscillator and the field can alter the state of the field so drastically that it invalidates the interpretation of the oscillator as a detector of the field \cite{LH07}.  Finite switch-on time also brings in out-of-equilibrium behavior. At the moment of being switched on the detector is at the ground state 
and is therefore out of equilibrium with the thermal or non-thermal bath of the field quanta. 

To connect to the more familiar equilibrium physics we show that near equilibrium and in the 
Markovian regime, which refers to the ultraweak coupling limit and/or the ultrahigh acceleration limit \cite{LH07},
the detector heats up analogously to the way classical bodies heat in fluids. The cooling-rate constant for Newton's law of cooling is calculated, providing a suggestive analogy to classical thermodynamics. Furthermore, and more important for the purpose of this paper, we discuss the range of validity of applying this type of reasoning for near-equilibrium condition to non-stationary situations, emphasizing that the non-thermal noise seen by the detector can lead to counter-intuitive late-time temperature readings if one naively makes assumptions about the instantaneous thermality of field based on the simple Unruh formula.

In Section \ref{sec:desdet}, we provide details of the detector model and the numerical strategy used to calculate the detector covariance matrix and its effective temperature. In Section \ref{sec:UADoutofequilib} we discuss the uniformly accelerated detector when its temperature is out of equilibrium with the thermal field. The subtle non-thermal effects that arise in the vacuum fluctuations when the motion is not uniform are discussed using the well-studied example of circular motion to explain this point, with numerical results provided in Appendix \ref{sec:circ}. In Section \ref{sec:numResults} we present our numerical results for the non-stationary oscillating detectors, followed by discussions in Section \ref{sec:discussion}. Finally our conclusion is given in Section \ref{sec:conclusion}.

\section{Description of the detector}\label{sec:desdet}
Consider an Unruh-DeWitt (UD) detector, whose internal degree of freedom is that of a harmonic oscillator $Q$, of mass $m_0$ and bare natural frequency $\Omega_0$, coupled with a massless scalar quantum field $\Phi$. The action of the combined system \cite{LH06} is given by:

\begin{eqnarray}
  S &=& -\int d^4 x \sqrt{-g} {1\over 2}\partial_\mu\Phi(x) \partial^\mu\Phi(x) + \nonumber\\ & &
  \int d\tau 
   \left\{ {m_0\over 2}\left[\left(\partial_\tau Q\right)^2
    -\Omega_{0}^2 Q^2\right]
    + \lambda_0\int d^4 x  Q(\tau)\Phi (x)\delta^4\left(x^{\mu}-z^{\mu}(\tau)\right)\right\},
  \label{Stot1}
\end{eqnarray}
where $g_{\mu\nu} = {\rm diag}(-1,1,1,1)$, $z^{\mu}$ is the worldline of the detector parametrized by the proper time $\tau$ and $\lambda_0$ is the coupling constant between the detector and field.

Assume the initial state (at $t=\tau=0$) of the combined system is a direct product of the ground state of the detector and the Minkowski vacuum of the massless scalar field. Since at most quadratic terms appear in the action, the state of the detector will remain  Gaussian throughout its evolution. Gaussian states are completely described by their covariance matrix, ${\rm v}_{ij}(\tau)=\langle \hat{R}_i(\tau),\hat{R}_j(\tau)\rangle\equiv {1\over 2}\langle \hat{R}_i(\tau)\hat{R}_j(\tau)+ \hat{R}_j(\tau)\hat{R}_i(\tau)\rangle$
and first order moments, $\langle \hat{R}_i\rangle$, where $\hat{R}_i=(\hat{Q}, \hat{P})$ and $\hat{P}= m_0 d\hat{Q}/d\tau$. For the assumed choice of initial state, the first order moments vanish and the covariance matrix separates into a sum of two parts \cite{LH06}:  ${\bf v}={\bf v}^{\rm a}+{\bf v}^{\rm v}$. The first part, ${\bf v}^{\rm a}$ ($= \langle \hat{R}_i(\tau),\hat{R}_j(\tau)\rangle_{\rm a}$ in \cite{LH06}),
accounts for the dissipation of zero-point energy initially in the detector.
It is independent of the motion of the detector when expressed as a function of the detector's proper time.
Explicitly, the ${\bf v}^{\rm a}$ matrix elements  are \cite{LH07}:
\begin{eqnarray}
{\rm v}^{\rm a}_{11}(\tau) &=&
  {\hbar \theta(\tau)\over 2\Omega^2\Omega_r m_0}e^{-2\gamma\tau}
  \left[\Omega_r^2 - \gamma^2\cos 2\Omega\tau +
  \gamma\Omega\sin 2\Omega\tau\right],\label{q2a}\\
{\rm v}^{\rm a}_{22}(\tau) &=&
  {\hbar \Omega_r m_0 \over 2\Omega^2}\theta(\tau)e^{-2\gamma\tau}\left[
  \Omega_r^2 -\gamma^2\cos 2\Omega\tau -\gamma\Omega\sin 2\Omega\tau\right],
  \label{dotQ2a}
\end{eqnarray}
and the off-diagonal elements can be calculated from ${\rm v}^{\rm a}_{12} = {\rm v}^{\rm a}_{21} = (m_0/2)\partial_\tau {\rm v}^{\rm a}_{11}(\tau)$,
where $\gamma \equiv \lambda_0^2/8\pi m_0$, $\Omega\equiv\sqrt{\Omega_r^2-\gamma^2}$ and $\Omega_r$
is the renormalized natural frequency of the detector.

On the other hand ${\bf v}^{\rm v}$ ($= \langle \hat{R}_i(\tau),\hat{R}_j(\tau)\rangle_{\rm v}$ in \cite{LH06}) includes the detector's response to vacuum fluctuations. It depends on the motion of the detector and can be obtained numerically
\cite{OLMH12} for any worldline by calculating the double integrals:
\begin{eqnarray}
  {\rm v}^{\rm v}_{11}=\langle \hat{Q}^2(\tau)\rangle_{\rm v} 
   &=& {\lambda_0^2\over m_0^2 \Omega^2} \lim_{\tau'\to\tau} \hspace{-1mm}Re
  \int_{0}^\tau \hspace{-2mm} d\tilde{\tau} \int_{0}^{\tau'} \hspace{-2mm} d\tilde{\tau}'
   K(\tau - \tilde{\tau}) K(\tau' - \tilde{\tau}')
   D^+(z^{\mu}(\tilde{\tau}), z^{\mu}(\tilde{\tau}')), \label{Q2int} \\
  {\rm v}^{\rm v}_{22}= \langle \hat{P}^2(\tau)\rangle_{\rm v} 
   &=& {\lambda_0^2\over \Omega^2} \lim_{\tau'\to\tau} Re
  \int_{0}^\tau d\tilde{\tau} \int_{0}^{\tau'} d\tilde{\tau}'
   \dot{K}(\tau - \tilde{\tau}) \dot{K}(\tau' - \tilde{\tau}')
   D^+(z^{\mu}(\tilde{\tau}), z^{\mu}(\tilde{\tau}')), \label{QPint}
\end{eqnarray}
and ${\rm v}^{\rm v}_{12}={\rm v}^{\rm v}_{21}= (m_0/2)\partial_\tau {\rm v}^{\rm v}_{11}$,
where $K(x)\equiv e^{-\gamma x}\sin\Omega x$ and $\dot{K}(\tau)=\partial_\tau K(\tau)$, and $D^+(z^{\mu}(\tilde{\tau}), z^{\mu}(\tilde{\tau}'))$ is the positive-frequency Wightman function of the massless scalar field along the worldline $z^\mu(\tau)$:
\begin{equation}
  D^+(z^\mu(\tau), z^\mu(\tau')) = {\hbar/(2\pi)^2\over \left| {\bf z}(\tau- i\epsilon/2)-{\bf z}(\tau'+i\epsilon/2)\right|^2
  - \left[ z^0(\tau- i\epsilon/2)-z^0(\tau'+i\epsilon/2) \right]^2},
\label{WFz}
\end{equation}
which encodes the quantum fluctuations of the Minkowski vacuum along the respective worldline and can be interpreted as a type of ``noise'' \cite{Sciama81, TAKAGI}. Notice that the Wightman function diverges in the coincidence limit:
\begin{equation}\label{Wsingstruc}
 \lim_{\tau\rightarrow\tau'}D^+(z^\mu(\tau), z^\mu(\tau')) = \lim_{\Delta \to 0}
 \frac{-\hbar/(2\pi)^2}{(\Delta-i\epsilon)^2\left[1+\frac{1}{12}a^2({\cal T})\Delta^2+\mathcal{O}(\Delta^4)\right]},
\end{equation}
where we have written $\tau \equiv {\cal T} + \Delta/2$, $\tau^\prime \equiv {\cal T} - \Delta/2$, and
$a({\cal T})\equiv \sqrt{a_\mu({\cal T}) a^\mu ({\cal T})}$ is the instantaneous proper acceleration of the detector at time ${\cal T}$. Since the $\tilde{\tau}$ and $\tilde{\tau}'$ integration variables in (\ref{Q2int})-(\ref{QPint}) can coincide inside the integration domain, special care is needed when numerically calculating these integrals \cite{OLMH12}. Equation (\ref{Wsingstruc}) shows that the singular behavior is independent of the worldline in question and in certain cases the integrals are exactly solvable.  Therefore, in the strategy we adopt, one first subtracts from the Wightman function of the worldline in question the Wightman function for the exactly solved worldline. Here we use the Uniformly Accelerated (UA)
detector \cite{LH06,LH07}:
\begin{equation}
  f_{a}\left(z^{\mu}(\tilde{\tau}), z^{\mu}(\tilde{\tau}')\right)
  \equiv D^+\left(z^{\mu}(\tilde{\tau}), z^{\mu}(\tilde{\tau}')\right)-
         D^+\left(z_{\rm UA}^{\mu(a)}(\tilde{\tau}), z_{\rm UA}^{\mu(a)}(\tilde{\tau}')\right), \label{fadef}
\end{equation}
where $z_{\rm UA}^{\mu(a)}$ the worldline of a UA detector with proper acceleration $a$.
Since this subtracted Wightman function is analytic, the resulting integrals with $f_a$ can be evaluated numerically. Then the analytic results for the UA detector correlators  \cite{LH06, LH07}, whose divergence is under control, are simply added back at the end of the calculation. For example,
\begin{equation}\label{eqn:Qtotalv}
  \langle \hat{Q}^2(\tau)\rangle_{\rm v} = \langle \hat{Q}^2(\tau)\rangle_{\rm v,UA}+\delta \langle \hat{Q}^2(\tau)\rangle_{\rm v},
\end{equation}
where the closed form expression for $\langle \hat{Q}^2(\tau)\rangle_{\rm v, UA}$ can be found in Eq.(A3) of \cite{LH07} and
\begin{equation}
\delta \langle \hat{Q}^2(\tau)\rangle_{\rm v} \equiv {\lambda_0^2\over m_0^2 \Omega^2} Re
  \int_{0}^\tau d\tilde{\tau} \int_{0}^{\tau} d\tilde{\tau}'
  K(\tau - \tilde{\tau}) K(\tau - \tilde{\tau}')  f_{a}\left(z^{\mu}(\tilde{\tau}), z^{\mu}(\tilde{\tau}')\right),
\label{dQ2v} 
\end{equation}
is evaluated numerically.

The bi-linear coupling in the last term of (\ref{Stot1}) results over time in 
multi-mode squeezings between the detector oscillator mode and the field modes.
The reduced state of the detector is obtained by tracing out the field modes in a multi-mode squeezed state which
then appears like a thermal state after its density matrix in the eigen-energy basis is diagonalized.
The effective temperature associated with the reduced state of the detector is evaluated in \cite{LH07} :
\begin{equation}
  T_{\rm eff}(\tau) = \left[{k_B\over \hbar\Omega_r}\ln\left({{\cal U(\tau)}+\hbar/2\over {\cal U(\tau)}-\hbar/2}\right)\right]^{-1},
\label{TeffDef}
\end{equation}
with ${\cal U}(\tau)\equiv \sqrt{\langle \hat{P}^2(\tau)\rangle\langle \hat{Q}^2(\tau)\rangle-\langle \hat{Q}(\tau),\hat{P}(\tau)\rangle^2}$
\footnote{$\langle \hat{Q},\hat{P}\rangle^2$ term is not present in Eq.(33) of Ref. \cite{LH07} because it goes to zero at late times.}.
In general, the effective temperature depends on all of the parameters in the model and the time of the evolution.

When the noise along the worldline is stationary (i.e., $D^{+}$ depends on $\Delta$ only) the power spectrum of the noise can be defined according to the Wiener-Khintchine theorem by:
\begin{equation}
{\cal F}(\omega)=\lim_{\epsilon\rightarrow 0} \int^{\infty}_{-\infty} d\Delta e^{-i \omega \Delta -\epsilon |\Delta|}D^{+}(\Delta)
\end{equation}
It is well known \cite{BD82} that the Wightman function for a massless scalar field in Minkowski vacuum along the UA detector worldline, 
$z^{\mu(a)}_{\rm UA}=(a^{-1}\sinh a\tau, 0,0, a^{-1}\cosh a\tau)$, is stationary:
\begin{equation}
  D^{+(a)}_{\rm UA}(\tau,\tau') \equiv 
  D^+\left(z_{\rm UA}^{\mu(a)}(\tau), z_{\rm UA}^{\mu(a)}(\tau')\right)  
  = -{\hbar\over (2\pi)^2} {a^2 \over 4 \sinh^2 {a\over 2}(\Delta-i\epsilon)} ,\label{WFUAD}
\end{equation}
and has an exactly thermal power spectrum
\begin{equation}
{\cal F}(\omega)=\frac{\hbar\omega}{2\pi}\frac{1}{e^{\hbar\omega/k_BT_U}-1},
\end{equation}
at the Unruh temperature (\ref{Unruhequation}).

Ideally, a good thermometer should not significantly disturb the temperature of the system it is measuring \cite{Berry}.  But as shown in \cite{LH07} when the detector (\ref{Stot1}) is accelerated uniformly, non-Markovian effects can result in late-time detector readings that are totally different from the Unruh temperature. In these regimes the backreaction of the detector strongly affects the field that it is measuring. When the interaction is turned on the system experiences a ``jolt" over very short time-scales of the order of the inverse UV cutoff frequency, which results in sudden entanglement generation between the detector and the modes of the field. Furthermore, the inclusion of backreaction self-consistently engenders memory effects in the dynamics. Because of these memory effects there is only a limited range of validity in which the model (\ref{Stot1}) can be considered to be a good thermometer for measuring the Unruh effect. Indeed, the effective temperature of the UA detector at late times is approximately the Unruh temperature only in the Markovian regime, which refers to the ultraweak coupling limit and/or the ultrahigh acceleration limit.

Fortunately, in realistic cases 
\footnote{Please note an erratum in \cite{LH07} in the heading and the text of section IV.A:
``$\Omega \Lambda_1\gg a, \gamma$''
should read 
$\gamma \Lambda_1\gg a, \Omega$''.}
the coupling is ultraweak and so the weak coupling limit applies \cite{LH07}. To briefly justify this claim, consider  $(\ref{Stot1})$ as an approximation to the interaction between a hydrogen-like atom (with approximately equally spaced energy levels) and the EM field.  The cut-off energy for this system will be the ionization energy of the outer-most electron $E_{\rm cutoff}\sim n \hbar \Omega_r$, (with $n$ equally spaced electronic levels). The dimensionless UV cutoff defined in \cite{LH07} is $\Lambda_1=\Lambda_0=-\ln{\hbar\Omega_r /E_{\rm cutoff}}\sim \mathcal{O}(1)$.   For hydrogen-like atoms the spontaneous emission rate of the transition from the first excited state to the ground state is $\hbar\gamma\sim10^{-7}$eV, and $\hbar\Omega_r\sim1$eV. Typically $a\sim \Omega_r/\hbar$ is the acceleration at which the Unruh effect becomes relevant. Comparing the parameters, we observe that this idealized atomic system is modeled by the action (\ref{Stot1}) in the ultraweak coupling regime $\gamma\Lambda_1\ll a,\Omega_r$. In this regime a UA detector has a late-time temperature consistent with the Unruh temperature \cite{LH07}.

\section{The uniformly accelerated detector out of equilibrium}\label{sec:UADoutofequilib}

\begin{figure}[h]
\includegraphics{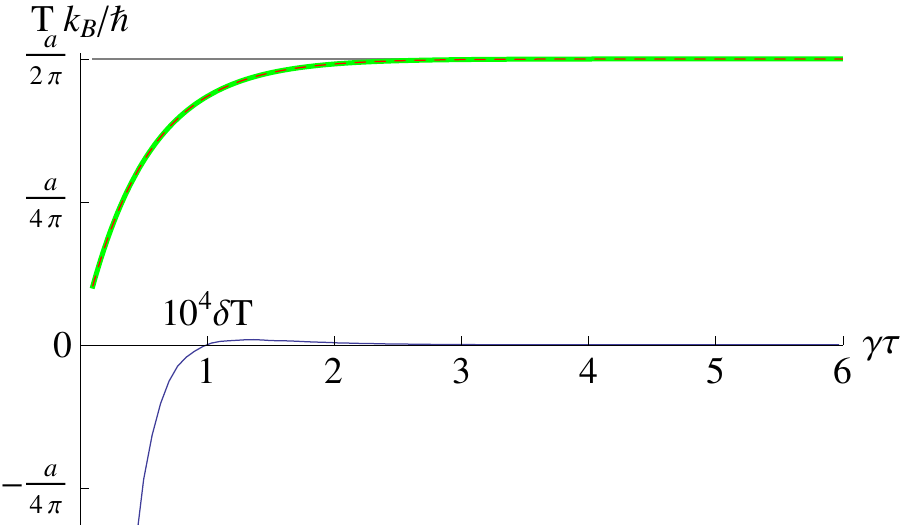}
\caption{(Color online) $a=100,
\gamma=10^{-3}, \Omega_r=1, \Lambda_0=\Lambda_1=3$. The green line is the temperature of the UA detector (grey is the
Unruh temperature), red dashed line is the solution to Newton's equation for $k=2\gamma$ and initial condition $T(\gamma^{-1})=T_{\rm UA}(\gamma^{-1})$. The blue line underneath shows the error between the two temperatures magnified by a factor of four orders of magnitude. }
\label{fig:NewtonUAD}
\end{figure}

Suppose a UA detector prepared in the ground state and far out of equilibrium has its coupling to the field 
turned on at some finite time. This induces a ``jolt'' which suddenly raises the detector's temperature from zero to a finite value. However, after a period of time of roughly the inverse coupling constant squared,  to good approximation the behavior of the evolution in the Markovian regime 
proceeds according to Newton's cooling equation.
To justify this claim, suppose that the temperature of the detector
near-equilibrium $|T_{\infty}-T|/T_{\infty}\ll 1$, obeys the evolution equation:
\begin{equation}\label{Newtoncooling}
\frac{dT}{d\tau}= k(T_{\infty}-T)+\mathcal{O}(T_{\infty}-T)^2.  
\end{equation}
i.e., Newton's law of cooling. If this were true of the UA detector, it would be possible to find a temperature independent value of the cooling rate, $k$. Using the analytic formulas in \cite{LH07} we find, in the ultraweak coupling limit (taking $\gamma$ to zero while keeping $\gamma \tau$ fixed), the simple formula:
\begin{equation}
k =\frac{1}{(T_{\infty}-T)}\frac{dT}{d\tau}=\left(2+\frac{\pi^2\Omega_r^2{\text {cosech} }{ (\tau^2 \gamma^2)}}{6 a^2}\right)\gamma +\mathcal{O}(\gamma^2).
\end{equation}
We see that at late times (near equilibrium), a fairly natural value of $k\sim 2\gamma$ emerges.  This approximate evolution is compared with the exact behavior of the UA detector in Fig.~\ref{fig:NewtonUAD}, showing very good agreement. While this correspondence to Newton's equation is only mathematical, it paints a very simple and intuitive picture for the thermalization process of Unruh-DeWitt detectors, especially when one compares it with the rather complicated  exact expression for the UA detector temperature \cite{LH07}.

\begin{figure}[h]
\includegraphics{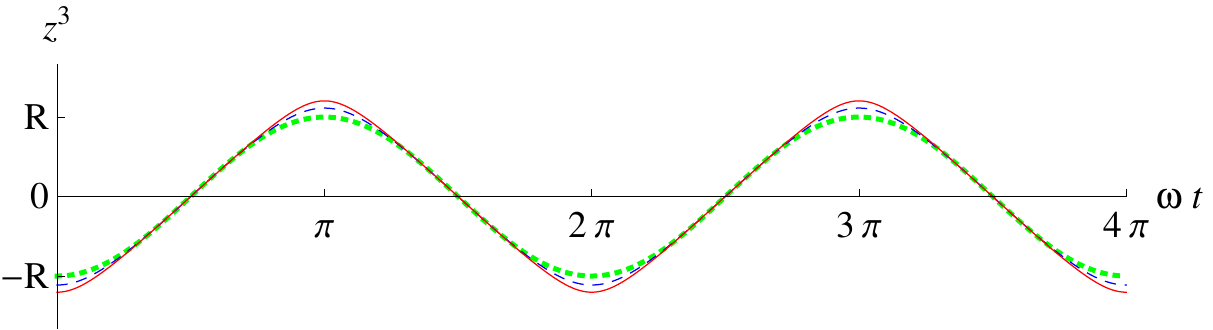}
\includegraphics{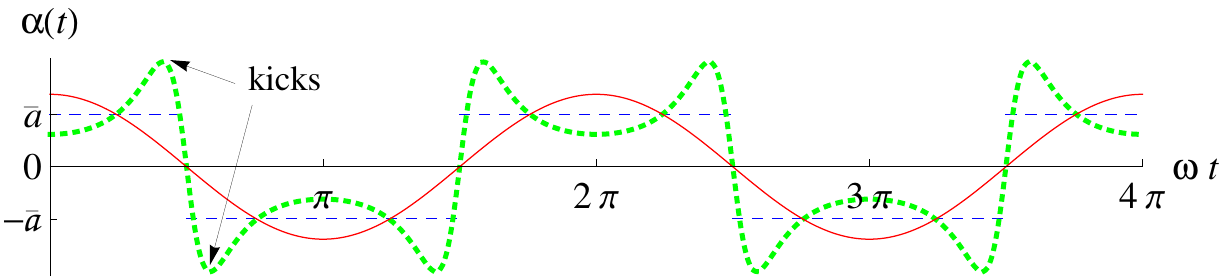}
\caption{(Color online) (Upper) Comparison of the oscillatory worldlines: $z^\mu_{\rm CT}$ red lined (\ref{zCT}),  $z^\mu_{\rm AUA}$ blue dashed (\ref{zAUAD}), and $z^\mu_{\rm SM}$ green dotted (\ref{zSM}). For each case the time-averaged proper acceleration  
and frequency are $\bar{a}=15$ and $\omega=10$ respectively. In the AUA case $a=\bar{a}=15$, in the other cases, $R(\bar{a},w)\approx 0.093$ 
and $a_0(\bar{a}, w)\approx 1.044$ 
have been obtained numerically. While the worldlines are nearly indistinguishable their accelerations are quite different: (Lower) Comparison of the directional proper accelerations $\alpha(t)$ ($a(t) = |\alpha(t)|$) for the three worldlines, for the same choice of parameters above. The double kicks in the SM acceleration profile become more enhanced in the ultra-relativistic regime. }
\label{fig:compSHMAUACT}
\end{figure}

Consider now applying this classical intuition to predict behaviour of  the temperature of an oscillating detector, which is by nature
always out of equilibrium with the apparent state of the field. For illustrative purposes the three kinds of oscillating worldlines
that we consider, and their corresponding directional proper accelerations, are compared in Fig.~\ref{fig:compSHMAUACT}.
The exact equations for these worldlines can be found in the next section.
In these figures the worldlines have been normalized such that they oscillate at equal coordinate frequency, $w$,
and have equal time-averaged proper accelerations, which we calculate by integrating the time-dependent proper accelerations, $a(\tau)$,
over a period $\tau \in {\cal P} = [\tau (t=0), \tau (t=2\pi/\omega)]$,
\begin{eqnarray}\label{timeavga}
  \bar{a} &\equiv& {\int_{\cal P}  a(\tau) d\tau \over \int_{\cal P} d\tau }.
\end{eqnarray}
Other mean values, e.g.  $a_{\rm rms} \sim \sqrt{(\int a^2 d\tau / \int d\tau)}$ could also be considered. However we will see in what follows that (\ref{timeavga}) is the most relevant average for our discussions.

One might be tempted to speculate that in the long-time limit the average temperature that the oscillating detector observes is given by Unruh's relation with the acceleration replaced by the time-averaged proper acceleration (\ref{timeavga}). 
Indeed this has been shown to be a good approximation for an very slowly varying acceleration with 
$\dot{a}/a^2\ll 1$ in \cite{KP10, BV12}.
To see why one might arrive at this, let us momentarily assume that the noise seen by the detector at each moment of time is instantaneously thermal and consistent with the Unruh temperature formula:
\begin{equation}\label{timetempassumption}
T(\tau)=\frac{\hbar a(\tau)}{2\pi k_B},
\end{equation}
where $a(\tau)$ is the instantaneous proper acceleration at time $\tau$. The proper time is used because the temperature is measured in the proper frame of the detector. According to this picture, the detector experiences a thermal bath with a time-dependent temperature. To answer how a detector behaves in such an environment we consider a classical analogy. Imagine a thermometer under a tap that is running water. The temperature of the water can be made to regularly oscillate by adjusting the hot-water faucet. At late times the reading on the thermometer must lie between the minimum and maximum temperatures of the water because it can always absorb or emit heat. One can model the situation using Newton's law of cooling:
$dT/dt =k (T_{\text{ext}}(t) -T),$
for some constant $k$, with time-dependent external bath temperature, $T_{\text{ext}}(t)$.
For any periodic $T_{\text{ext}}(t) $ there will be a steady state solution. At late times this solution must satisfy $\overline{dT/dt}=0$,
where over-line represents the time-averaged value, otherwise the temperature would vary from one period to the next.
It then follows from Newton's equation that $\overline{T}(t)=\overline{T}_{\text{ext}}(t)$ in the steady state regime.
In other words, at late times the average reading on the detector equals the average temperature of the time-dependent surroundings if the time-scale of the response of the thermometer is much longer than the period of the oscillation of temperature.

Since we know that the near-equilibrium evolution of the UA detector also follows the Newton equation in the Markovian regime,
by analogy one expects that the temperature recorded by an oscillating detector satisfies
\begin{equation}\label{Naiveprediction}
\overline{T}_{\text{eff}}(\infty)= \frac{\hbar \bar{a}}{2\pi k_B}.
\end{equation}

We will see in the next section that even in the Markovian 
regime where the UA detector approaches the Unruh temperature at late times, such formula does not always work. 
The reason is because assumption (\ref{timetempassumption}) is not exactly correct.
This can already be seen in the well-studied case of circular motion.
To see this one only needs to compare the two-point correlation functions, i.e., the quantum ``noise'' of the field,
experienced by the detector. The worldline of a Circularly Moving (CM) detector is:
\begin{equation}
  z^\mu_{\circ}(\tau) =(\Gamma \tau, (\Gamma v/\omega)\sin\omega \tau, (\Gamma v/\omega)\cos\omega\tau,0),
\label{wdln}
\end{equation} 
where $\omega$ is the angular frequency of the circular motion, $v$ is the speed of the detector in Minkowski time and
$\Gamma \equiv 1/\sqrt{1-v^2}$. The proper acceleration, $a_{\circ} =  \Gamma v\omega$ is constant in time.
Expanding the positive-frequency Wightman function for the CM detector in terms of $\Delta = \tau-\tau'$, one obtains
\begin{eqnarray}\label{WFcirc}
  && \hspace{-5mm} D^+_{\circ}(\tau,\tau') = {\hbar\over (2\pi)^2}
  \left[ \left( {2\Gamma v\over \omega}\sin{\omega\Delta\over 2}\right)^2-\left(\Gamma\Delta\right)^2\right]^{-1} \nonumber\\
  &=& {\hbar\over (2\pi)^2} \left\{ \Delta^2
    \left[ -1 -{a_{\circ}^2\over 12}\Delta^2 + {a_{\circ}^2\omega^2\over 360}\Delta^4 -{a_{\circ}^2 \omega^4\over 20160}\Delta^6+
    O(\Delta^8)\right] \right\}^{-1}
  \nonumber\\ &=& {\hbar\over (2\pi)^2}\left[ -{1\over \Delta^2} + {a_{\circ}^2\over 12} -
  {a_{\circ}^4\over 720}\left( 3+{2\over v^2}\right)\Delta^2+{a_{\circ}^6\over 60480}\left(10 + {22\over v^2}+{3\over v^4} \right) \Delta^4
  + O(\Delta^6) \right]
\end{eqnarray}
with $\epsilon$ suppressed.
On the other hand, expanding the Wightman function for the UA detector (\ref{WFUAD}) with the same proper acceleration we have
\begin{eqnarray}
  D^{+(a_\circ)}_{\rm UA}(\tau,\tau') 
  &=& {\hbar\over (2\pi)^2} \left\{ \Delta^2
    \left[ -1 - {a_{\circ}^2\over 12}\Delta^2 -{a_{\circ}^4\over 360}\Delta^4 -{a_{\circ}^6\over 20160}\Delta^6 +
    O(\Delta^8)\right] \right\}^{-1} \nonumber\\
  &=& {\hbar\over (2\pi)^2}\left[ -{1\over \Delta^2} + {a_{\circ}^2\over 12} - {a_{\circ}^4\over 240}\Delta^2 +
  {a_{\circ}^6\over 6048}\Delta^4 + O(\Delta^6)\right].
\label{WFUADSer}
\end{eqnarray}
Although the first two terms (up to $O(\Delta^0)$) are equal, the higher order corrections differ, and thus the field noise statistics are not identical in these two cases
(note that $0\le v <1$, though apparently $D^+_{\circ}$ approaches $D^+_{\rm UA}$ when $v \to \infty$). 
This point was already made by Bell and Leinaas \cite{BL83}, and indeed as they found, the effective temperature in the circular case has a frequency dependent spectrum at a temperature slightly different from the Unruh temperature. 
A detailed analysis has been given by Unruh \cite{Unr98}.
In Appendix \ref{sec:circ} we provide 
more numerical results of the late-time effective temperature for the detector (\ref{Stot1}) in circular motion.

Kinematically the difference between the linear Unruh effect and the circular effect is rooted in the qualitative difference between linear acceleration and angular acceleration. Even on simple dimensional grounds, there is only one parameter in the linear case, the proper acceleration, whereas there are two in the circular case, the angular acceleration and the radius of the orbit (or equivalently, the speed).
In the linear case, the speed 
of the particle asymptotically approaches the speed of light, entailing the formation of an event horizon. In the circular case, the direction of velocity changes but its magnitude remains constant and there is no event horizon.
At a more fundamental level, the difference between the linear and circular Unruh effects reflects the deep divide between equilibrium and non-equilibrium conditions. From the kinematical viewpoint the circular case displays non-equilibrium (albeit steady state) quantum field statistics that are more general than the linear uniform acceleration case.

The lesson to be learned here is that simply knowing that a system is accelerating at an instant of time is not enough to conclude that the field statistics are instantaneously thermal at the Unruh temperature as in equation (\ref{timetempassumption}). Other details of the motion, such as the velocity in the case of circular motion, also affect the state of the field as it appears to the moving observer.  These will affect the temperature registered by a detector and can lead to counter-intuitive results, some of which will be discussed in the next section.

\section{Oscillating Detectors: Results}\label{sec:numResults}

In this section we present our results for the detector (\ref{Stot1}) moving along the three oscillating worldlines shown in Fig.~\ref{fig:compSHMAUACT}. As mentioned, in order for the accelerating oscillator to function like a thermometer of the field it is important that we work in the Markovian regime \cite{LH07}.
While the ultraweak coupling regime is likely to be the one relevant to experiments, working at these very weak coupling regimes can be intensive numerically because of the extremely long evolution time that is required to reach steady state.
In order to obtain an understanding of the qualitative features of realistic detectors we instead consider another regime in which the UA detector observes an Unruh temperature: the ultrahigh acceleration limit $a\gg \gamma \Lambda_1,\Omega$ \cite{LH07}. Uniformly accelerated detectors in this regime will also reproduce the characteristic Unruh behavior at late times.

\subsection{SM worldline: sinusoidal oscillations}

The first case we consider is Sinusoidal Motion (SM), which is a projection of the detector in circular motion onto the
$z^0$-$z^3$ subspace. The worldline is:
\begin{equation} \label{zSM}
  z^{\mu}_{\rm SM}(t)=\Big(t,0,0, -R \cos{\omega t}\Big),
\end{equation}
where $R$ is the oscillation amplitude, and $\omega$ is the oscillation frequency in coordinate time. The constraint $R\omega<1$ is necessary in order for the motion to remain time-like. The proper time can be explicitly solved as a function of the coordinate time using the elliptic integral of the second kind: $\tau=\omega^{-1} E(\omega t, (R\omega)^2)$, which can be inverted numerically to find $t(\tau)$.  The directional proper acceleration is $\alpha^{}_{\rm SM}(t) = 
Rw^2 \cos{\omega t}(1-(Rw)^2 \sin^2{\omega t})^{-3/2}$, the proper acceleration $a^{}_{\rm SM}(t)= |\alpha^{}_{\rm SM}(t)|$,
and the period of oscillations is $t_p=2\pi/\omega$ and $\tau_p= \omega^{-1} E(2\pi, (R\omega)^2)$ in the coordinate time and proper-time respectively.
The time-averaged proper acceleration (over one period of oscillation) is:
\begin{eqnarray}
  \bar{a} &=& {\omega \tanh^{-1}R \omega \over E\left(R^2\omega^2\right)}.
\end{eqnarray}

\begin{figure}[h]
\includegraphics[width=4.8cm]{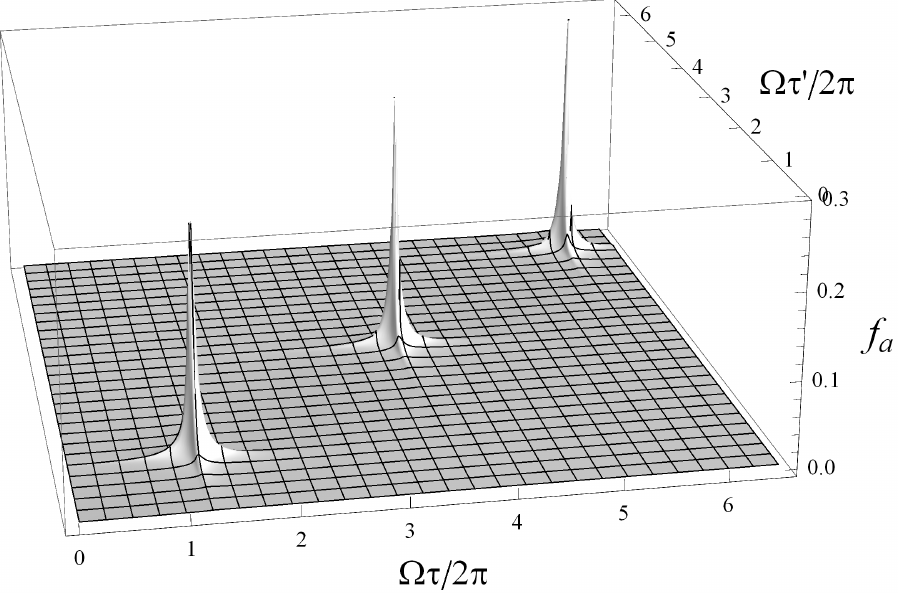}
\includegraphics[width=4.8cm]{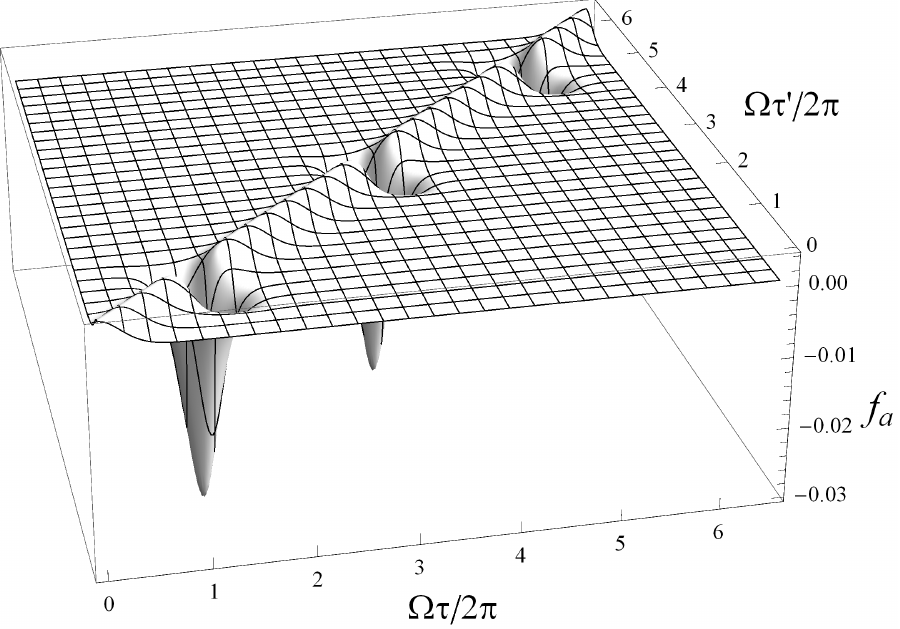}
\includegraphics[width=4.8cm]{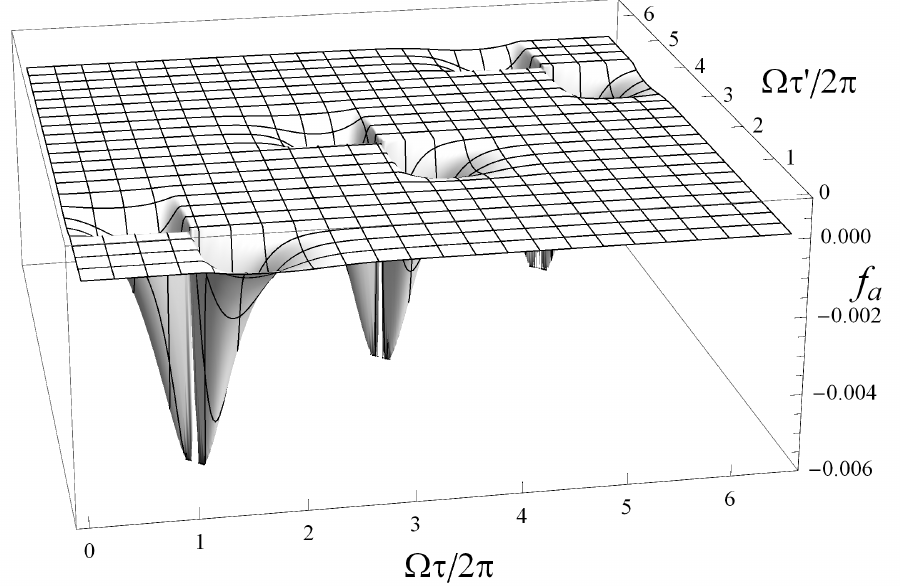}
\caption{
(Left) The behavior of $a^{}_{\rm SM}(\tau)$ below (\ref{zSM}) creates double-peaks of $f_{\bar{a}}$ around
$t(\tau)\approx t(\tau') \approx (2k+1)/(2\pi \omega)$, $k=1,2,3,\ldots$. 
They give positive ``kicks" to the subtracted two-point correlators  and the effective temperature, 
whose mean values eventually saturate to some constants but
the nonadiabatic oscillations on top of them are always there. 
In contrast, both the $f_{\bar{a}}$ in the CT (middle) and the AUA (right) cases have periodic negative dips around
$t(\tau)\approx t(\tau') \approx (2k+1)/(2\pi \omega)$, which give some kicked down behavior in the evolution of the 
subtracted correlators  and the effective temperatures.
}
\label{alphafa}
\end{figure}

As can be seen from the lower plot in Fig. \ref{fig:compSHMAUACT}, the acceleration profile for the SM worldline develops extra peaks due to relativistic dilation effects that are largely amplified by the factor $\Gamma^3 = (1-v^2)^{-3/2}=(1-(Rw)^2\sin^2{\omega t})^{-3/2}$ around $t(\tau)\approx (2k +1)/(2\pi\omega)$, $k=1,2,3,\ldots$, when the speed of the detector is large. 
This creates periodic positive double-peaks in the function $f_{\bar{a}} \equiv D^+\Big(z^{\mu}_{SM}(\tilde{\tau}), z^{\mu}_{SM}(\tilde{\tau}')\Big)
-D^+\Big(z_{\rm UA}^{\mu(\bar{a})}(\tilde{\tau}), z_{\rm UA}^{\mu(\bar{a})}(\tilde{\tau}')\Big)$ defined
in Eq. (\ref{fadef}) with the maximum value
\begin{equation}
   f_{\bar{a}}^{\rm max} = {1\over (2\pi)^2}\left[ {4\over 27}\omega^2 \cosh^4 {\bar{a}\tau_p\over 2} - {\bar{a}^2\over 12}\right]
\label{fmax}
\end{equation}
along $\tau \approx \tau'$ (see Fig. \ref{alphafa} (left)),
which gives the periodic ``kicks" to the subtracted v-parts of the two-point correlators. 

\begin{figure}[h]
\includegraphics[width=4.8cm]{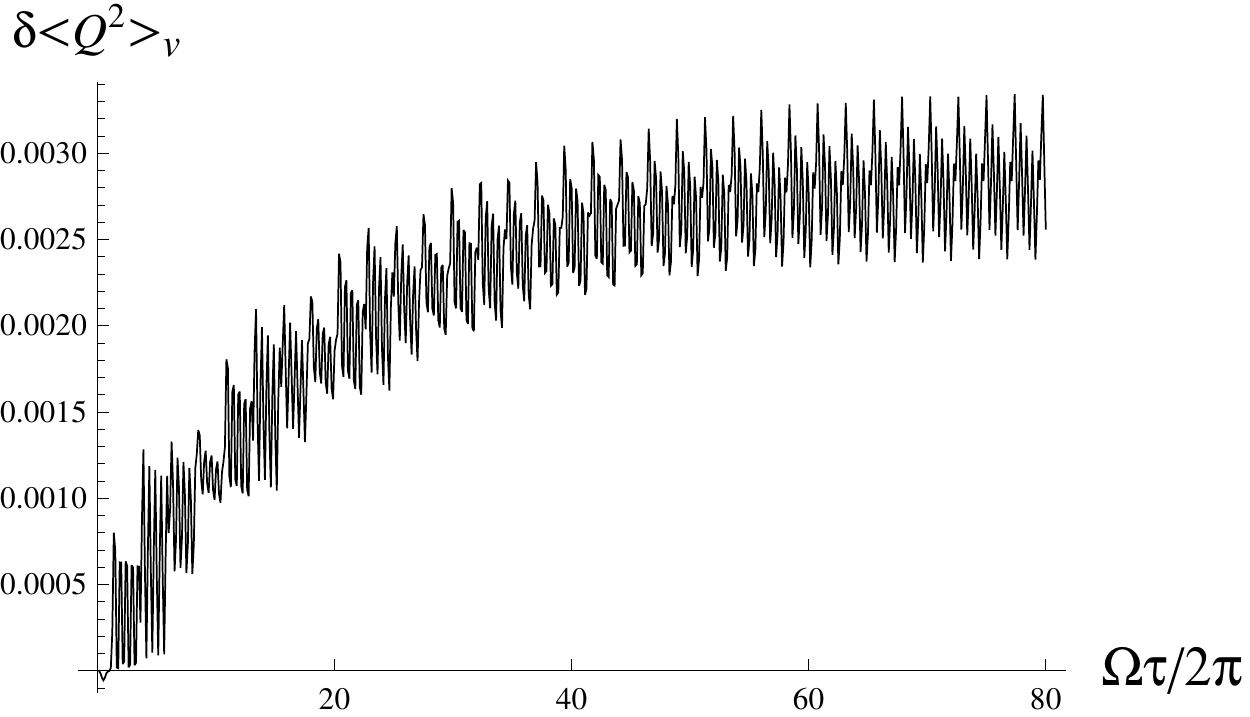} 
\includegraphics[width=4.8cm]{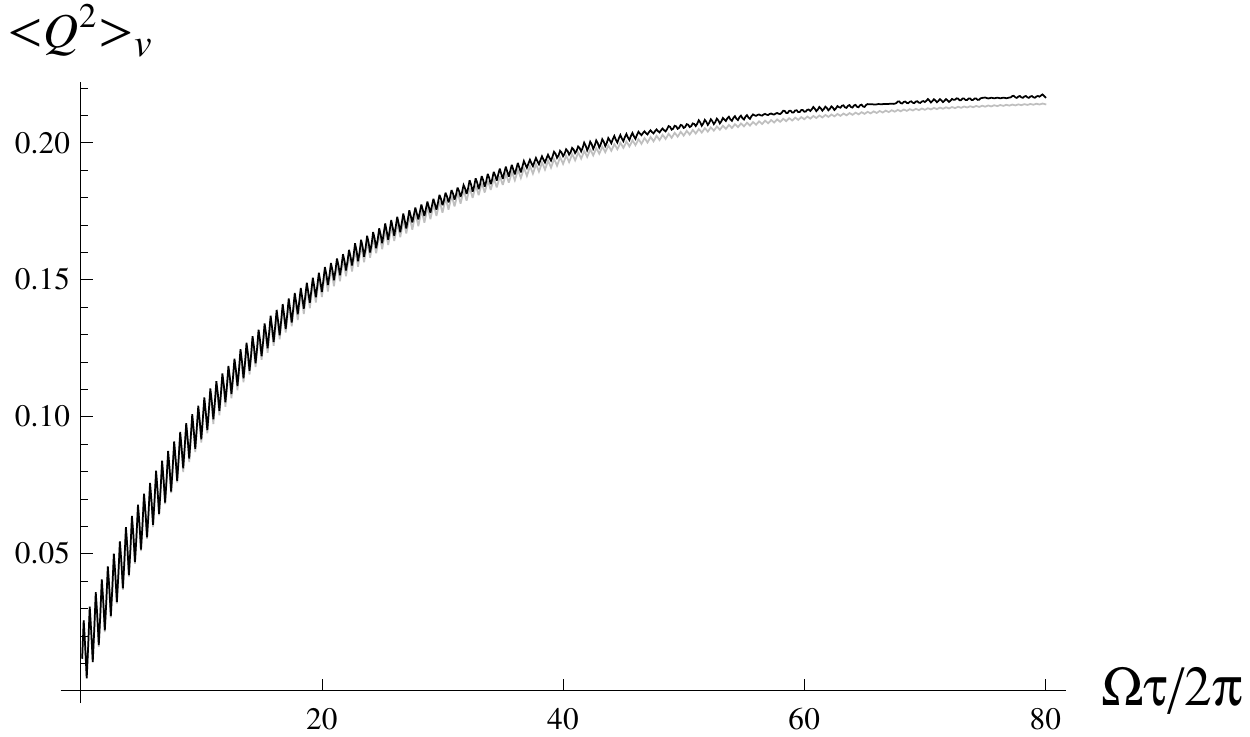} 
\includegraphics[width=4.8cm]{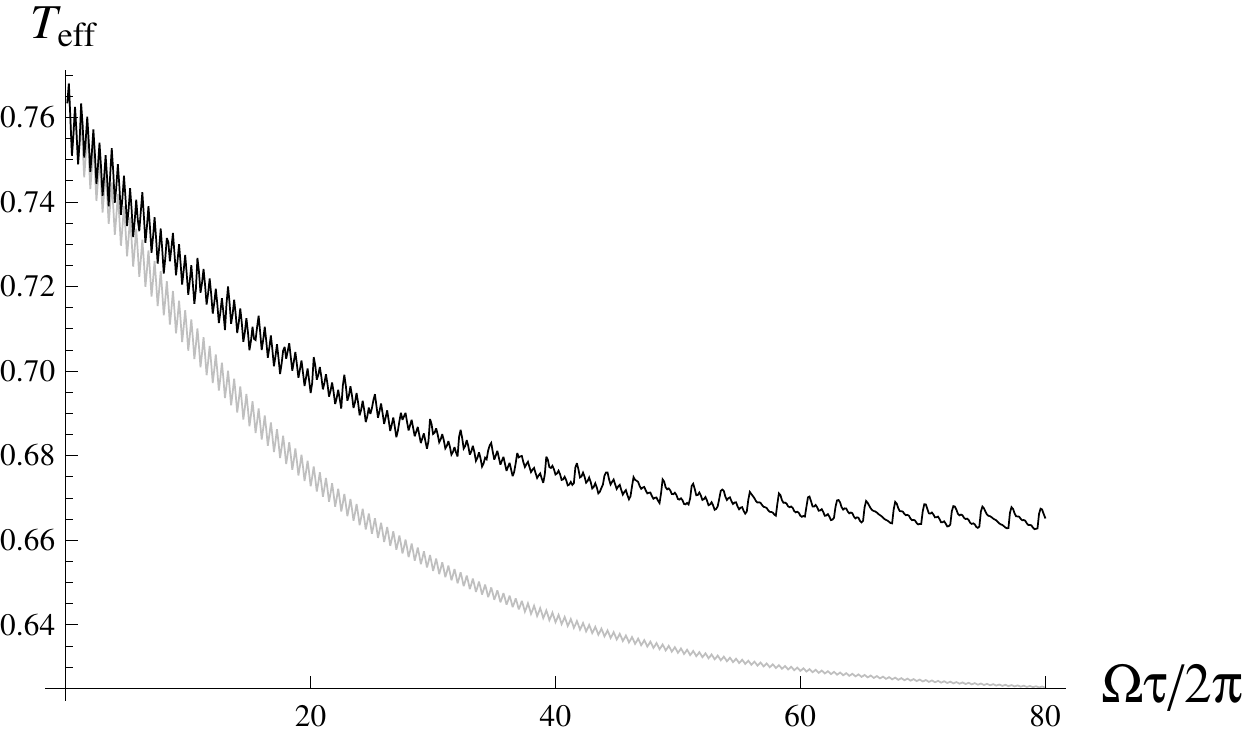}
\caption{
Numerical results for $\delta\langle \hat{Q}^2\rangle_{\rm v}$ (left, given in (\ref{dQ2v})),
$\langle \hat{Q}^2\rangle_{\rm v}$ (middle), and the effective temperature $T_{\rm eff}$ 
(right, given in Eqs.(\ref{eqn:Qtotalv}-\ref{TeffDef})) in the SM case (dark lines)
compared with those for the UA detector (grey lines) with the same $\bar{a}$.
In the middle plot the grey line is almost indistinguishable from the dark line: 
The difference between them is below $O(\gamma)$ at late times. 
Here $\gamma=0.01$, $\Omega=2.3$,
$\hbar=m_0=1$, $\tau_0=0$, $\Lambda_0=\Lambda_1=20$,
$\tau_p \approx 12.957$, 
$\omega \approx 0.314$, and $\bar{a} \approx 0.926$.
}\label{SHMvResult}
\end{figure}

We illustrate an example for the evolution of the correlators and the effective temperature of the UD detector in SM motion in Fig. \ref{SHMvResult}. 
One can see that the mean values of these quantities eventually settle down to constants, while the effects of the kicks on top of 
the mean values always persist. 
In particular, the sawtooth-like structure of the evolution curve for the effective temperature in Fig. \ref{SHMvResult} (right) reflects 
the effect of these periodic ``kicks" which kick the temperature up at regular intervals.
In between those kicks there are small ``nonadiabatic oscillations" with period $\tau_p$, which is due to the large rate of change of the subtracted Wightman function in the $\tau\tau'$-plane \cite{OLMH12}.

The numerical results for the mean values of the late-time temperature (averaged over a period of oscillation) of a detector following this worldline with $\omega=20$ and $\bar{a}$ from $5$ up to $15$, are shown in Fig.~\ref{TC} (right), where the maximum speed reached is $0.755$ when $\bar{a}=15$. One can see that at high accelerations the temperature becomes lower than what one might naively expect from the Unruh formula, Eq. (\ref{Naiveprediction}).

\subsection{CT worldline: Worldline of a charge in standing wave of intense laser}

To identify the effects with and without the positive ``kicks", below we consider an alternative case 
inspired by the experiment proposed by Chen and Tajima (CT) \cite{CT99}.
A particle, of mass $m$ and charge $e$, placed at one of the magnetic nodes of an EM standing wave,
of amplitude $E_0$ and frequency $\omega$, follows the worldline (without considering radiation reaction):
\begin{equation}\label{zCT}
 z^\mu_{\rm CT}(t) = \left( t, 0,0, -{1\over \omega}\sin^{-1}{2a_0\cos\omega t\over\sqrt{1+4a_0^2}}\right),
\end{equation}
where $a^{}_0 \equiv e E_0/m\omega$. The directional proper acceleration is $ \alpha^{}_{\rm CT}(t) = 2 a^{}_0 \omega \cos \omega t$.
The proper time of the detector $\tau$ is related to the coordinate time $t$ by $ \tau(t)  = \omega^{-1}F(\omega t, -4a_0^2)$, where $F(\phi, m)$ is the elliptic integral of the first kind. The inverse function $t(\tau)$ can be obtained numerically. This worldline oscillates with a period $t_p= 2\pi/\omega$ or a proper time period of $\tau_p= \omega^{-1}F(2 \pi, -4a_0^2)$. The time-averaged proper acceleration reads
\begin{equation}
  \bar{a} = {\omega \sinh^{-1}2a_0\over F\left( \pi/2, -4a_0^2\right)}.
\end{equation}
At low accelerations and non-relativistic velocities, $z_{\rm CT}\sim z_{\rm SM}$.

\begin{figure}[h]
\includegraphics[width=4.8cm]{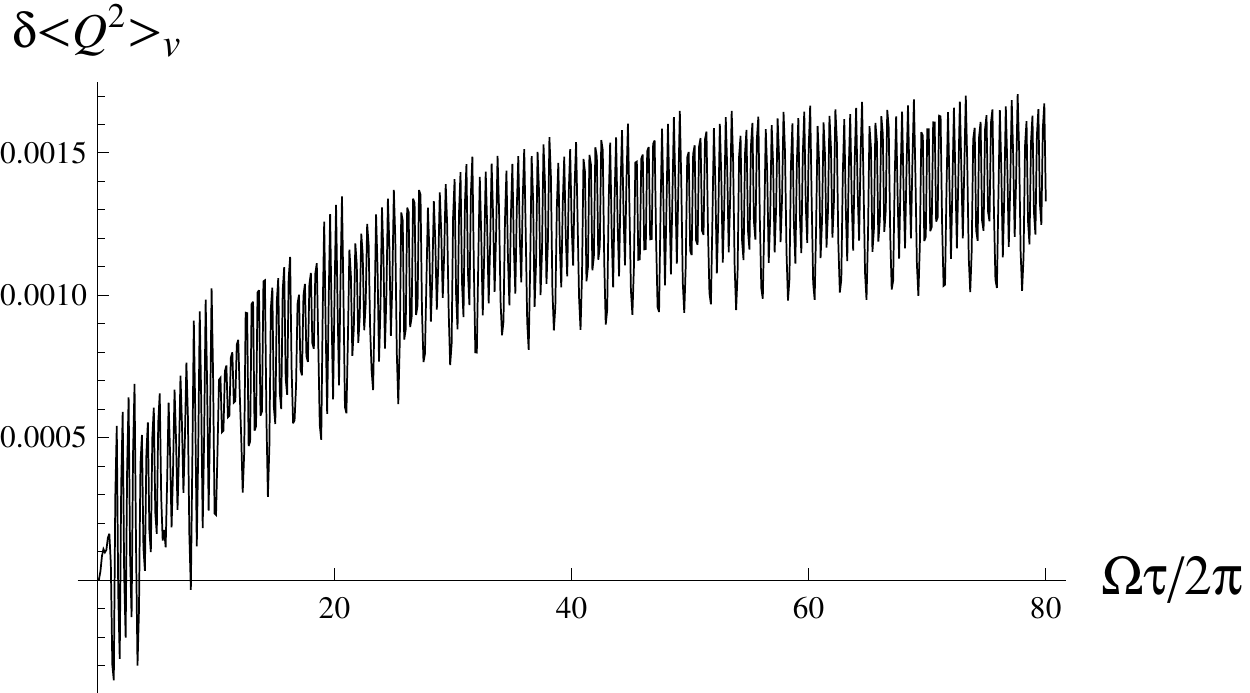} 
\includegraphics[width=4.8cm]{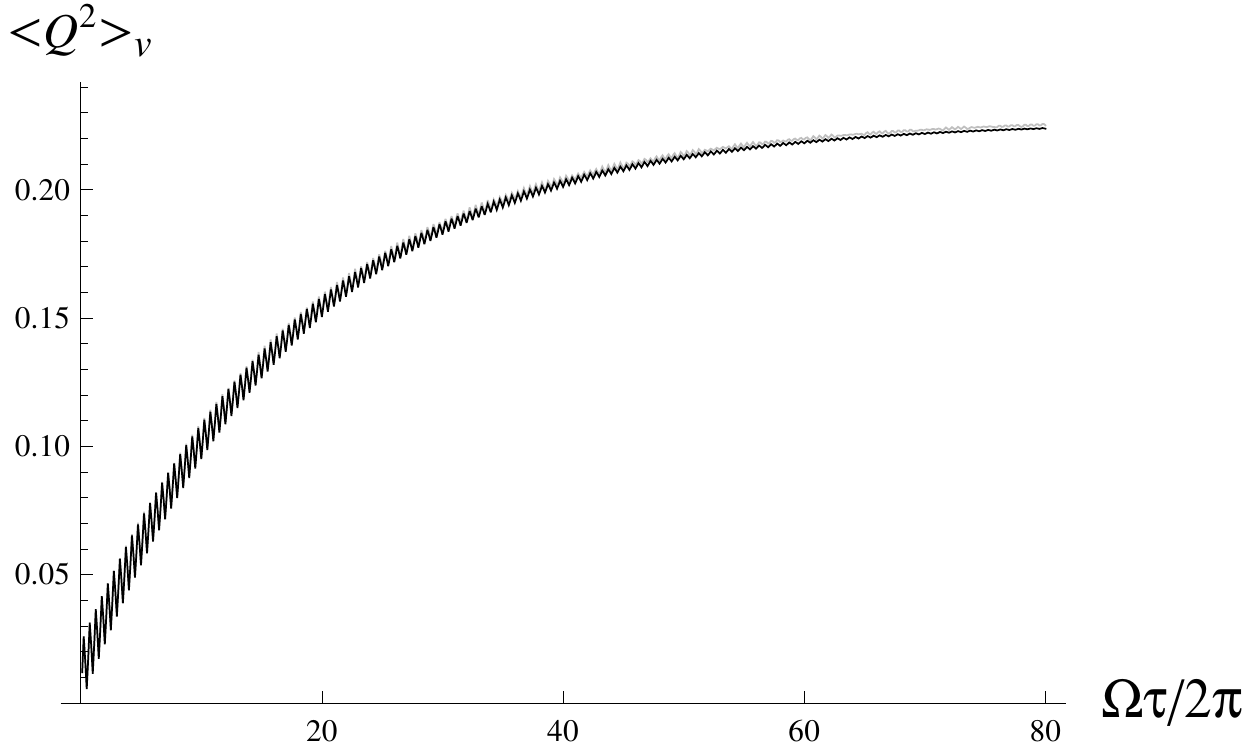} 
\includegraphics[width=4.8cm]{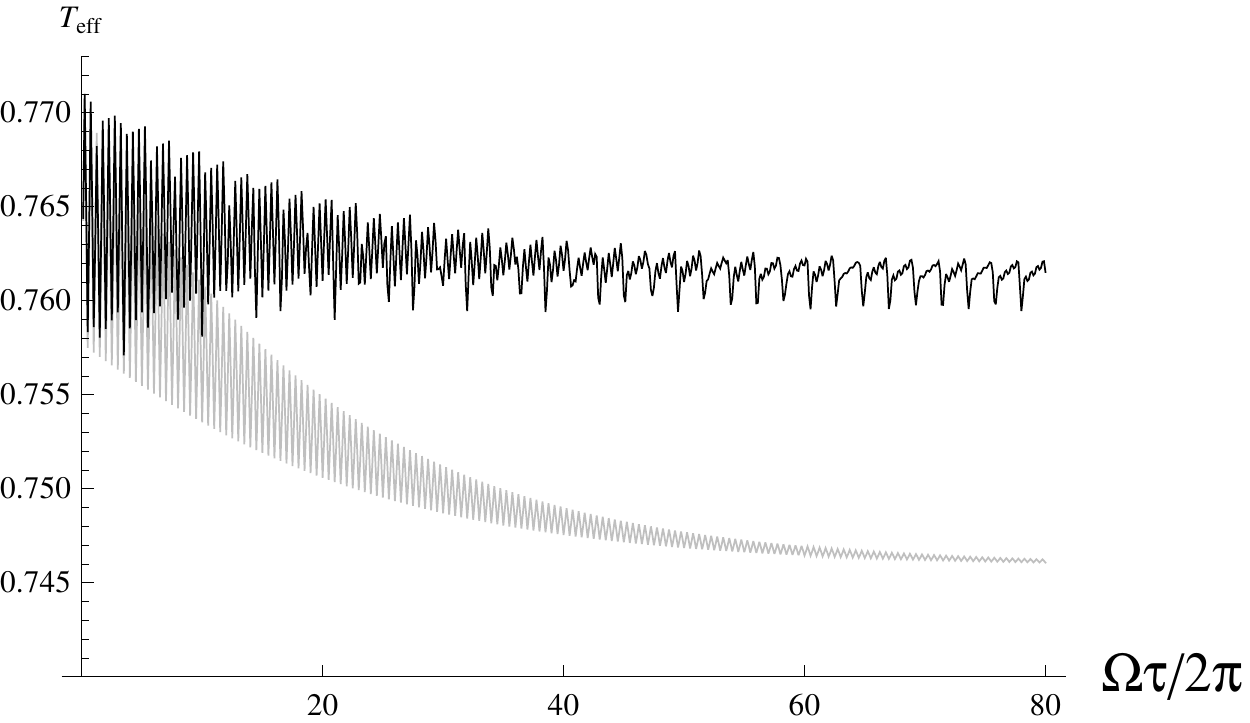}
\caption{
Numerical results of $\delta\langle Q^2\rangle_{\rm v}$ (left),  $\langle \hat{Q}^2 \rangle_{\rm v}$ 
(middle), and $T_{\rm eff}$ (right) for the CT detector (dark lines)
compared with the results for the UA detector (grey lines) with the same $\bar{a}$. 
Here $\omega=0.0001$, $a_0=20000$, $\bar{a}\approx 3.769$, and $\tau_p \approx 11.983$.
Other parameters are the same as those in Fig. \ref{SHMvResult}.
}\label{CTQ2Teff}
\end{figure}

An example for the evolution of the correlators and the effective temperature in the CT case is shown in Fig. \ref{CTQ2Teff}.
The behaviors of these quantities are quite similar to those in Fig. \ref{SHMvResult} 
in the SM case, but the periodic upward kicks in the SM case are now replaced by downward kicks due to the periodic dips of the proper acceleration around $t(\tau) \approx (2k+1)/(2\pi \omega)$, $k=1,2,3,\ldots$ (cf. Fig. \ref{fig:compSHMAUACT}).
Again at late times the effective temperature of the CT detector oscillates about a constant mean value.

We have numerically calculated the 
mean values of the late-time effective temperatures
of the CT detector at two different frequencies $\omega=1$ and $\omega=20$.
As can be seen from Fig.~\ref{TC}, the late-time behavior of the effective temperature of the detector
against $\bar{a}$ is different in these two cases.
For $\omega=1$ the temperature agrees quite well with the naive application of the Unruh relation (\ref{Naiveprediction}),
however for $\omega=20$, again, at high accelerations the temperature becomes lower than the Unruh temperature.

\subsection{AUA worldline: alternating uniform acceleration regime}

One should be aware of the complication in the CT and SM cases, where part of the motion around $t(\tau) = (2k+1)\pi/\omega$
($k=1,2,3,\ldots$) when $a \approx 0$ may produce non-Markovian effects even in the weak coupling limit \cite{LH07}.
To suppress this effect, below we consider the case with the proper acceleration
of the detector is constant except at the isolated moments $\tau = \tau_p (2k+1)/4$,
$k=1,2,3,\ldots$, when the proper acceleration is zero:
\begin{equation}
  a^\mu_{\rm AUA}(\tau)= \left(a \sinh a\left[\tau- {n\tau_p\over 2}\right], 0, 0, (-1)^{n}a \cosh a\left[\tau- {n\tau_p\over 2}\right]\right).
\end{equation}
Here $n(\tau)\equiv {\rm floor}\left( {2\tau\over\tau_p} +{1\over 2}\right)$,  and ${\rm floor}(x)$
gives the largest integer less than or equal to $x$. Then we have $\bar{a}=a$.
We call this the Alternating Uniform Acceleration (AUA) worldline:
\begin{eqnarray}
  z^\mu_{\rm AUA}(\tau) &=& \left( \frac{1}{a}\Big[\sinh a\left(\tau-\frac{n\tau_p}{2}\right)+2n\sinh {\frac{a\tau_p}{4}}\Big],0,0,\right.
  \nonumber\\ & &\qquad \left.{(-1)^n\over a}\Big[\cosh a\left(\tau-\frac{n\tau_p}{2}\right)+ \left\{(-1)^n-1\right\} \cosh {\frac{a\tau_p}{4}}\Big]\right).
\label{zAUAD}
\end{eqnarray}
Note that as $a_0$ increases with $\omega$ fixed, the CT-worldline converges to the AUA, so
the correlators and effective temperatures of both cases will converge in this high acceleration limit, too.
The direction of acceleration alternates in periods of proper time equal to $\tau_p/2$, with total oscillation period of $\tau_p$. Both $z^\mu_{\rm AUA}$ and its four velocity are continuous everywhere. The worldline oscillation period in coordinate time is $t_p= 4a^{-1}\sinh{a \tau_p/4} $ and its frequency is $\omega=2\pi/t_p$.

\begin{figure}[h]
\includegraphics[width=4.8cm]{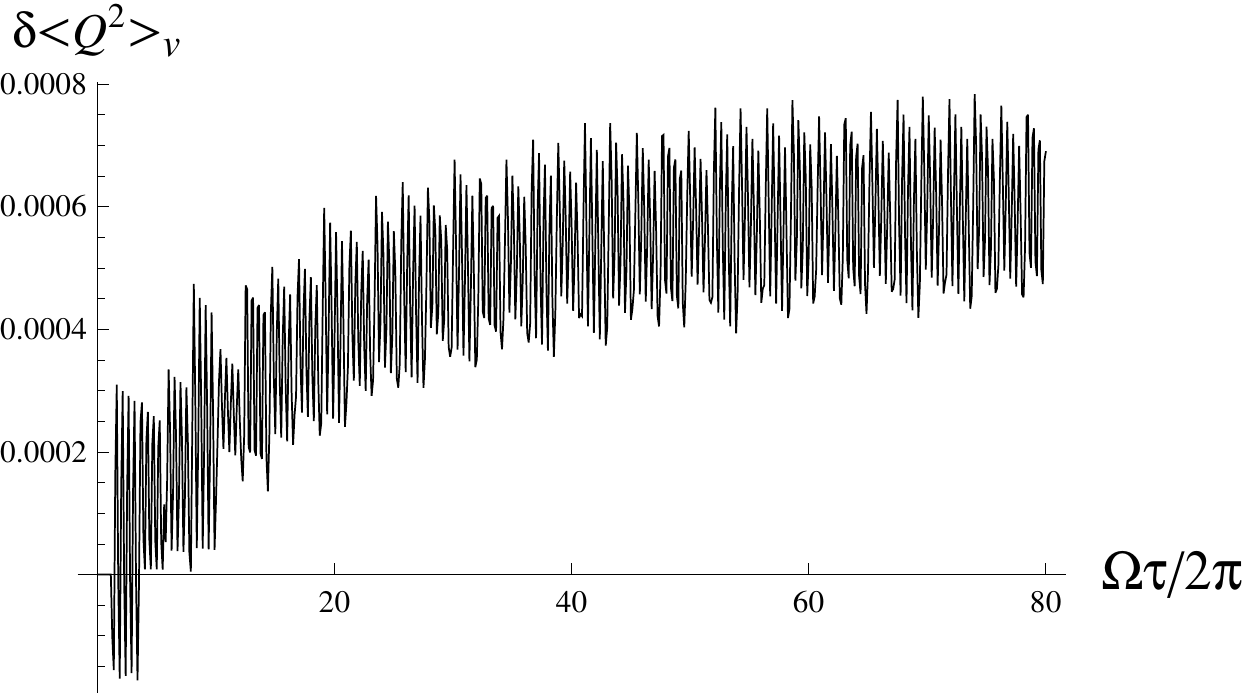} 
\includegraphics[width=4.8cm]{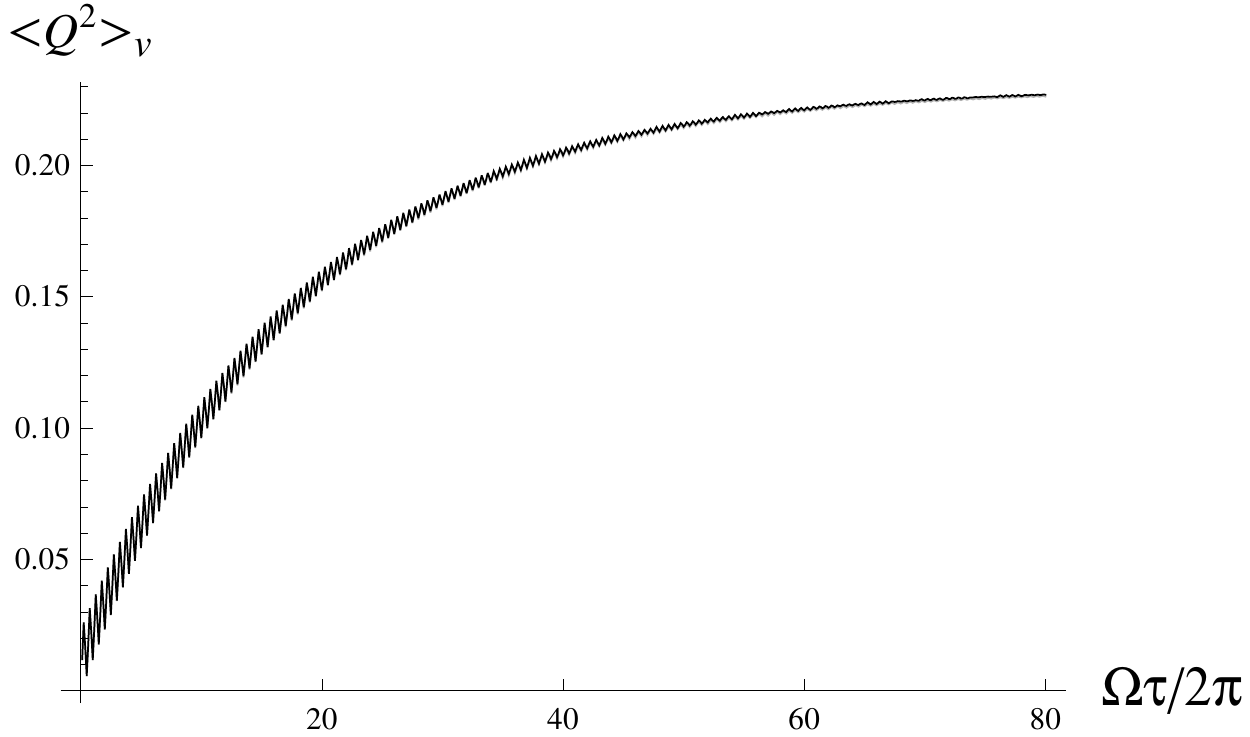} 
\includegraphics[width=4.8cm]{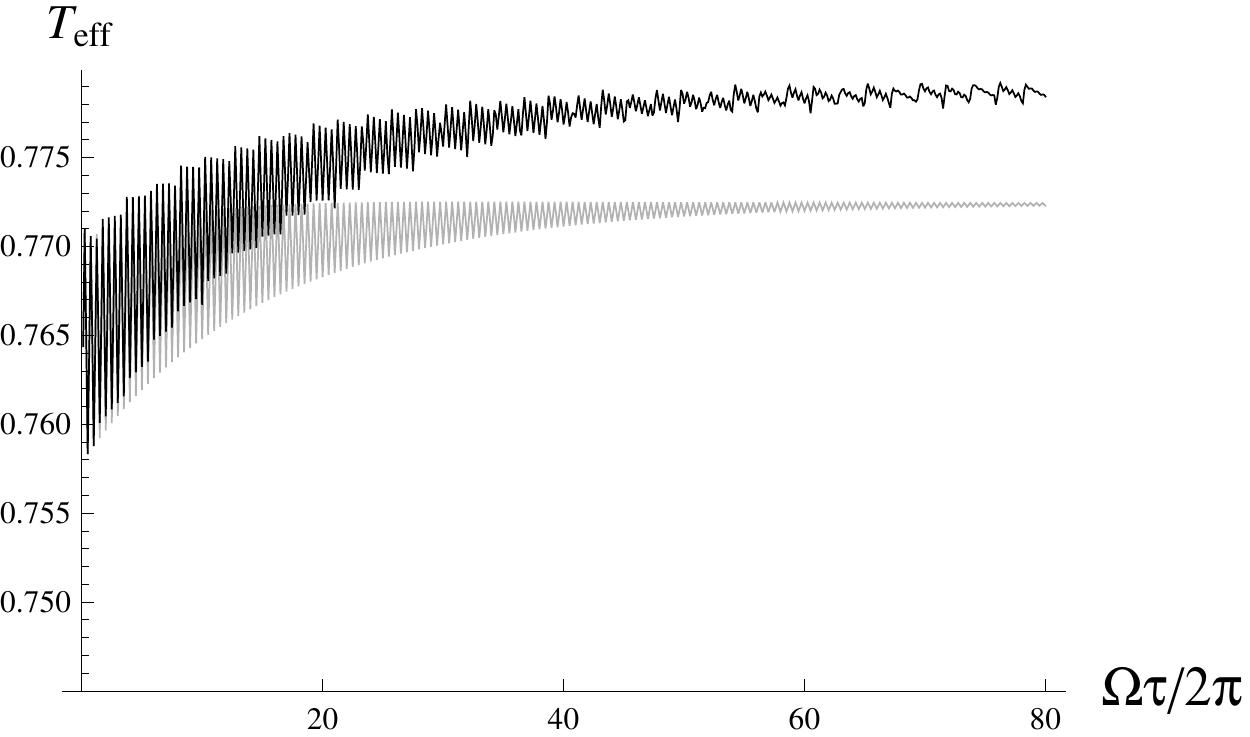}
\caption{
Numerical results of $\delta\langle Q^2\rangle_{\rm v}$ (left), $\langle \hat{Q}^2 \rangle_{\rm v}$ 
(middle), and $T_{\rm eff}$ (right) for the AUA detector (dark lines)
compared with the results for the UA detector (grey lines) with the same $\bar{a}$. 
Here $\tau_p=12$, $\bar{a}=a=4$, other parameters are the same as those in Fig. \ref{SHMvResult}.
Note that here the ratio of the period $\tau_p$ of the alternating acceleration and the natural period of
the detector $2\pi/\Omega$ is not a rational number.}
\label{Q2vResult}
\end{figure}

In Fig. \ref{Q2vResult} we give an example for the results 
in the AUA case when $a\sim O(\Omega) \gtrsim 2\pi/\tau_p$. 
We found that the late-time values of $\delta\langle \hat{Q}^2\rangle_{\rm v}$ and $\delta\langle \hat{P}^2\rangle_{\rm v}$ 
are both positive here, yielding a positive correction to the effective temperature of a UA detector in this parameter range. 
Compared with Figs. \ref{SHMvResult} and \ref{CTQ2Teff}, 
one can see that $T_{\rm eff}$ in the AUA case has the periodic ``kicked-down" behavior 
similar to the ones in the CT case, though the magnitude is smaller here. 
This can be explained by the periodic dips in the middle and the right plots of Fig. \ref{alphafa}.
Note that, while the AUA subtracted Wightman function $f_a(\tau, \tau')$ is constant almost everywhere along the 
line $\tau = \tau'$ in Fig. \ref{alphafa} (right), the dips are significant in the blocks with $|n(\tau)-n(\tau')|= 1$ or larger, 
namely, when $z^\mu_{\rm AUA}(\tau)$ and $z^\mu_{\rm AUA}(\tau')$ are {\it not} in the same UA piece of the AUA worldline. 

\begin{figure}[h]
\includegraphics[width=7cm]{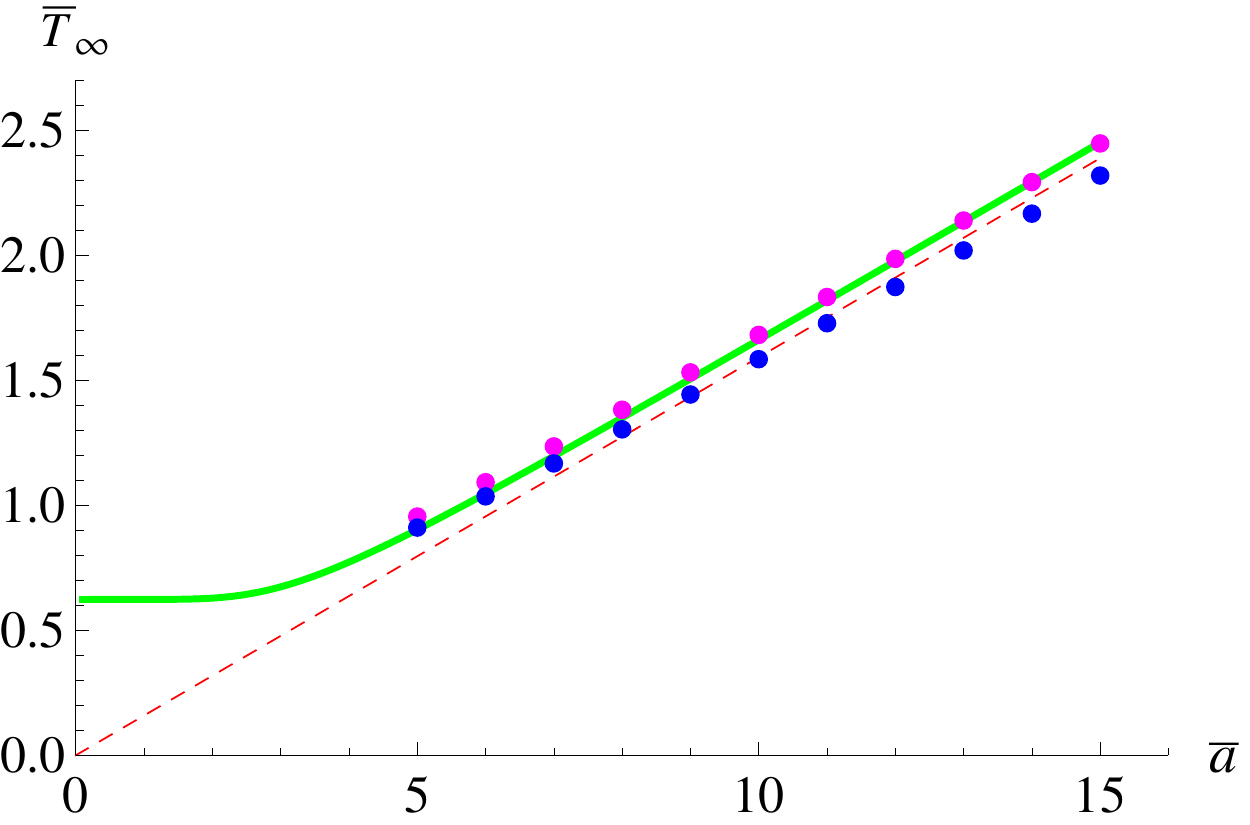}
\hspace{2mm}
\includegraphics[width=7cm]{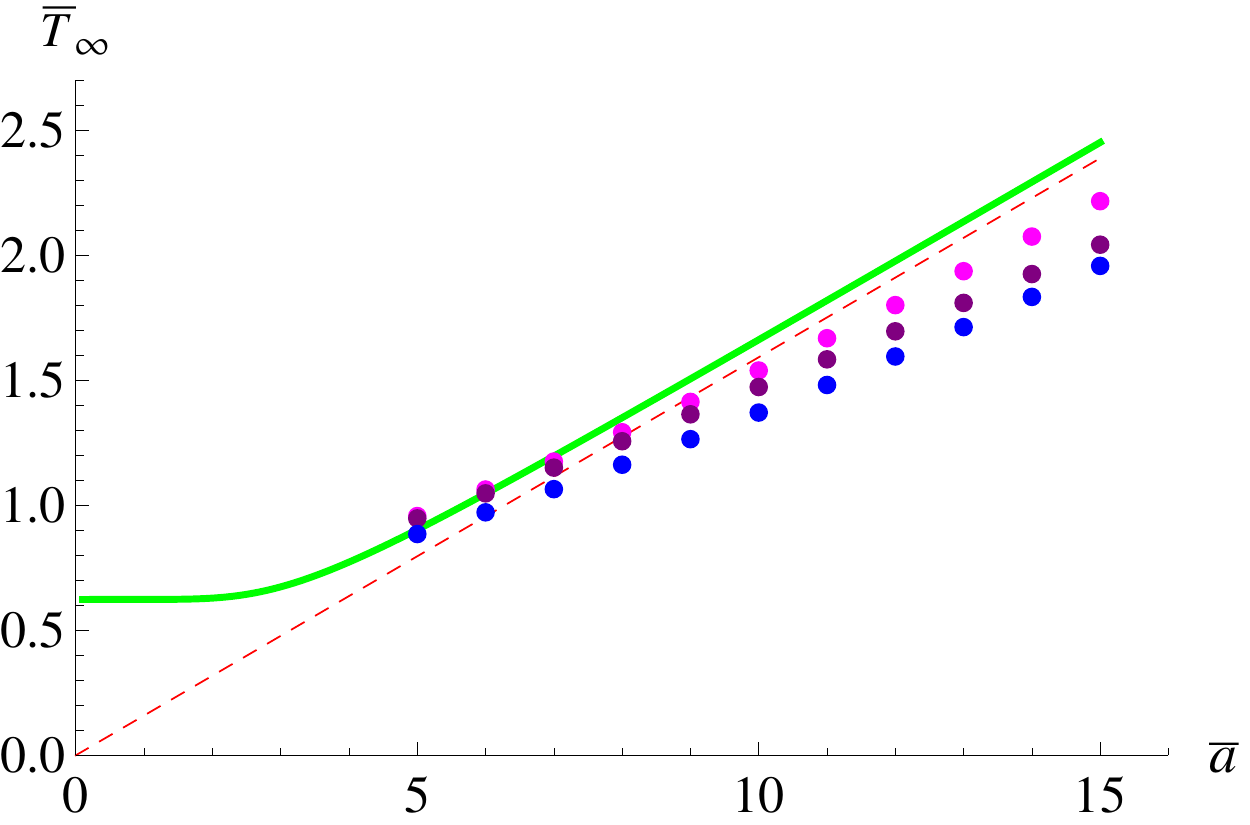}
\caption{Comparison of the average late-time temperatures, $\overline{T}_{\infty}$, of the numerically computed data points 
for the CT-worldline (magenta), SM-worldline (purple, right plot only) and AUA-worldline (blue) against 
the UA detector (green solid) and Unruh temperature (red dashed).
Large double kicks in the subtracted Wightman function for the SM case prevent us from numerically accessing the parameter space when $\omega=1$.   
The amplitude of the oscillations in the temperature at late times  
are much smaller than the size of the drawn data points. 
$\bar{a}$ is the time-averaged proper acceleration (\ref{timeavga}). Here $\omega=1$ (left) 
and $\omega=20$ (right); in both cases the parameters  $\gamma=0.01$, $\Omega=2.3$, $\Lambda_0=\Lambda_1=20$ have been chosen.
If the coupling strength $\gamma$ gets smaller, then $\overline{T}_{\infty}$ will be even lower, mainly because of the 
$\gamma \Lambda_1$ term in $\langle P^2 \rangle$ for the UA detector \cite{LH07}.
}
\label{TC}
\end{figure}

When one increases the value of the proper acceleration $a$ such that $a \gg \Omega$, 
the corrections to $\delta\langle \hat{Q}^2\rangle_{\rm v}$ and $\delta\langle \hat{P}^2\rangle_{\rm v}$, 
for the AUA detector become negative, and the averaged effective temperature at late times 
becomes lower than the Unruh temperature, as shown in Fig. \ref{TC}. 
This is consistent with the results for the detectors in the SM and CT cases in the same limit. 

As the oscillation frequency $\omega$ is lowered,
the period of each piece of uniform acceleration, $\tau_p/2$, gets longer, and one expects that  
the effective temperature of the AUA detector would more closely resemble the one for the UA detector. 
Indeed, the data points with $\omega=1$ (left plot in Fig. \ref{TC}) are closer to the results for the UA detectors
compared with the data points with $\omega=20$ (right).

\section{Discussion}\label{sec:discussion}
\label{disc}

\subsection{Comparison to Circular Motion and Uniform Linear Acceleration}
\label{compareCM}
The plots in Fig.~\ref{TC} show that the temperature experienced by an oscillating detector is much richer than one would expect from the naive formula (\ref{timetempassumption}).  In particular, one observes a transition from $\overline{T}_{\infty}(\bar{a}) >T_U$ to $\overline{T}_{\infty}(\bar{a}) <T_U$ as $\bar{a}$ is increased. 
In fact, it is already present in circular motion as reported in \cite{BL83,Unr98}. 

\begin{figure}[h]
\includegraphics[width=7cm]{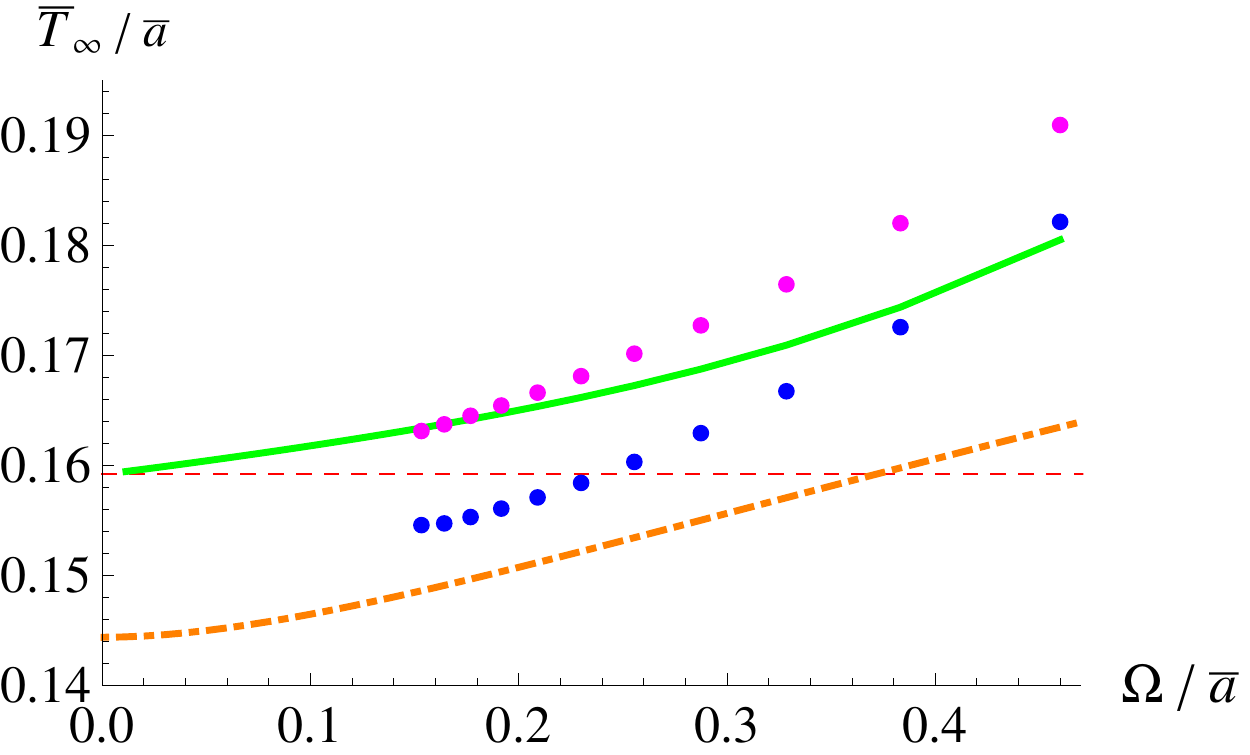}
\hspace{2mm}
\includegraphics[width=7cm]{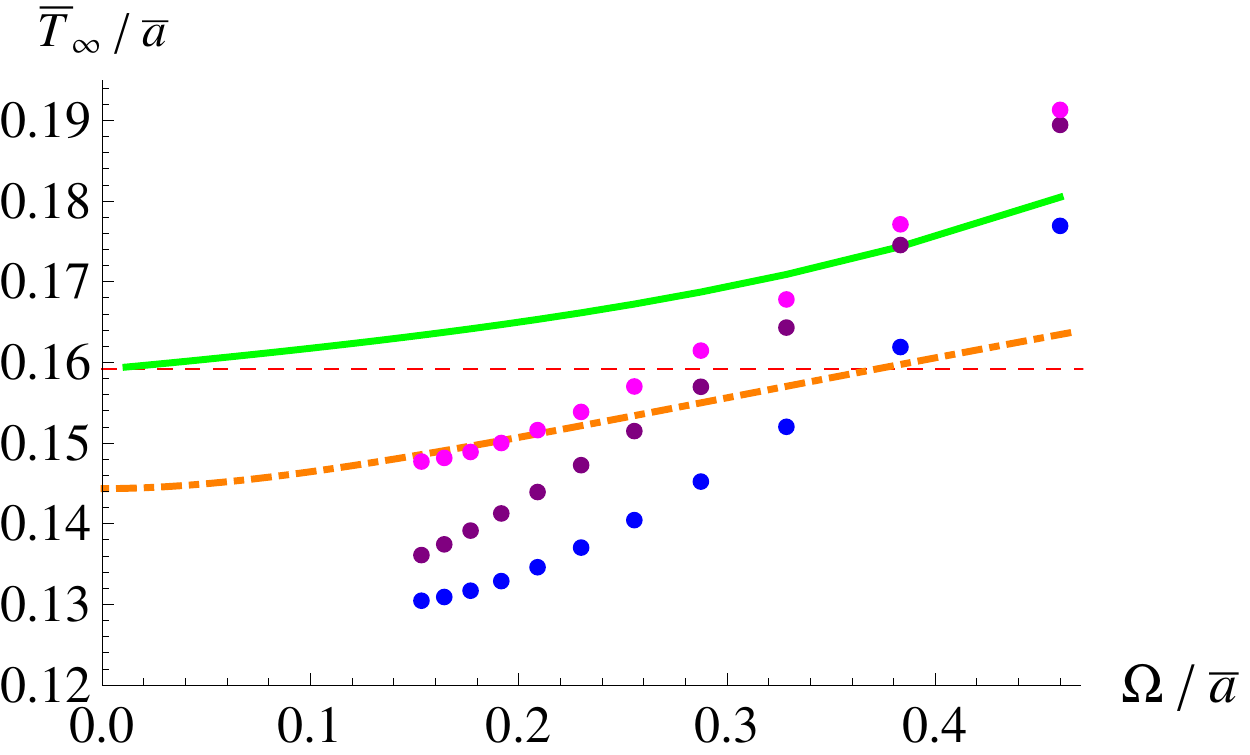}
\caption{Comparison of $\overline{T}_{\infty}/\bar{a}$ versus the normalized natural frequency of the detector, $\Omega/\bar{a}$ for the same data as in Fig.~\ref{TC}. Again we have $\omega=1$ in the left plot and $\omega=20$ in the right. Now this is compared to the known circular temperature formula (Eq. (12) in \cite{Unr98} with the $\omega$ there replaced by $\Omega$ in this paper) for circular motion (orange line). The green line is that of the UA detector which approaches the Unruh temperature formula (red dashed line) only in the large acceleration or weak coupling limit. 
We see that, though the lines for different oscillating worldlines are different, 
the oscillating worldlines and the circularly moving worldlines have similar transitional behaviour across the red dashed line.
However, in the ultrahigh acceleration limit ($\bar{a} \gg \Omega, \gamma$) the value of $\overline{T}_{\infty}/\bar{a}$ 
for each oscillating worldline depends on $\omega$, in contrast to the circularly moving detectors. 
When $\omega=20$, our results indicate that $\overline{T}_{\rm eff} \approx 0.13 \bar{a}$ for the AUA worldline 
and $\overline{T}_{\rm eff} \approx 0.14 \bar{a}$ for the CT worldline when $\bar{a}\gg \Omega$.
}
\label{TvsOmega}
\end{figure}

In Fig.~\ref{TvsOmega} we compare our numerical data points to a similar plot that was produced for circular motion previously \cite{BL83,Unr98}.
In these plots the behaviour of the effective temperature of the oscillating detector more closely resembles that of the circularly moving detector than the UA detector because their Wightman functions (\ref{WFz}) are more similar at large $\Delta = \tau-\tau'$. In the denominator of both the circularly moving and oscillating Wightman functions, the value of $|{\bf z}(\tau) - {\bf z}(\tau')|^2$ is always less than $(c t_p)^2$  (in CT, this is the wavelength of the driving lightfield $\sim 10^{-7}$m). As $|\Delta|$ grows larger than a few periods of the oscillations, the $[z^0(\tau) - z^0(\tau')]^2$ term dominates 
such that the oscillatory and CM Wightman functions go like $D^+ \sim -(\Gamma\Delta)^{-2}$. 
In contrast, $|{\bf z}(\tau) - {\bf z}(\tau')|^2$ for the UA detector is unbounded and grows with $[z^0(\tau) - z^0(\tau')]^2$,
so that the UA Wightman function goes like $D^+_{\rm UA}\sim -a^2 e^{-a\Delta}$ as $\Delta$ grows large.

\begin{figure}[h]
\includegraphics[width=7cm]{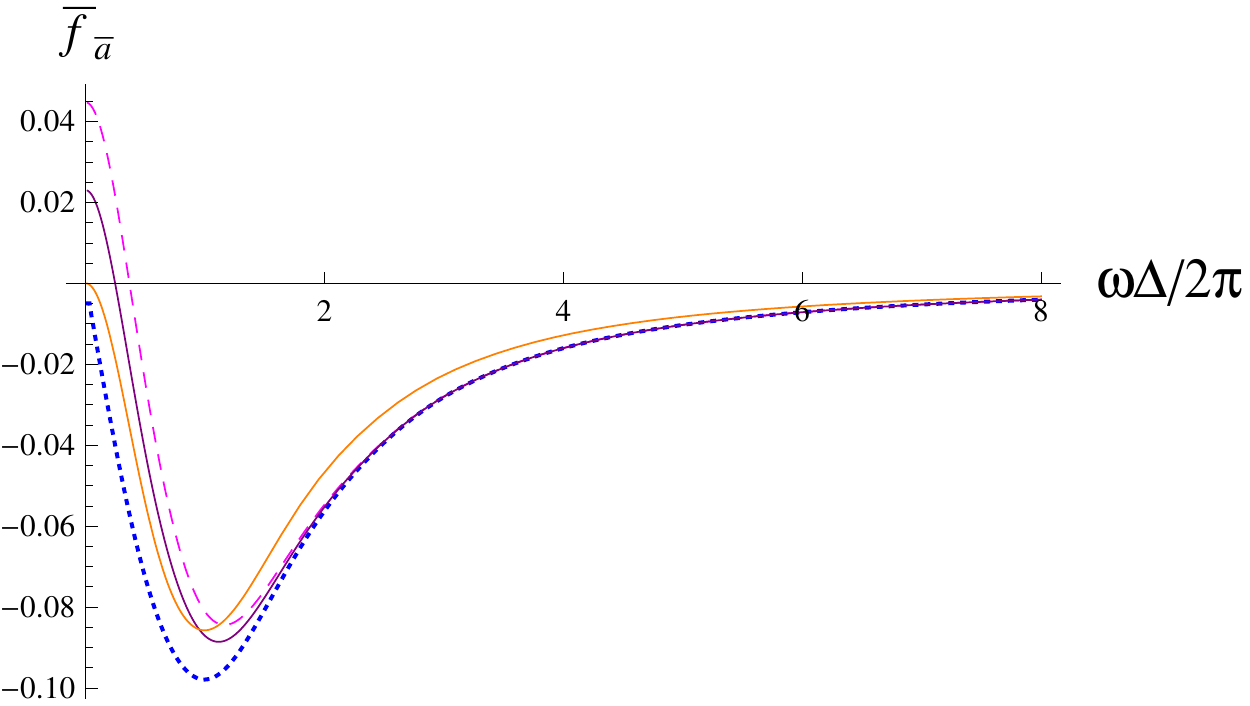}
\caption{(Color Online) Subtracted Wightman functions, $f_{\bar{a}}(\tau, \tau') = D^+ - D^+_{\rm UA}$ defined in (\ref{fadef}),  for the SM (purple), CT (magenta dashed) and AUA (blue dotted), circular (orange) worldlines averaged in the direction of ${\cal T}\equiv(\tau+\tau')/2$ over a period $\tau_p$ with $\bar{a}=a_\circ=10$ and $\omega=20$. The similarity suggests that the oscillating detectors behave more like circularly moving detectors than UA detectors when $\Omega\ll \omega$, $\bar{a}$.}
\label{Avgf}
\end{figure}

In Fig.~\ref{Avgf} we compare the subtracted Wightman functions $f_{\bar{a}}(\tau, \tau')$ for the SM, CT and AUA worldlines averaged in the direction of ${\cal T}\equiv(\tau+\tau')/2$ for one period $\tau_p$. We find that their shapes are also similar to that of the Wightman function of the detector in circular motion, though when $\Delta$ goes large, $f_{\bar{a}} \sim -\left( {2\omega\over \pi \bar{a}}\sinh^{-1}{\pi \bar{a}\over 2\omega}\right)^2\hbar /(2\pi \Delta)^2$ for the SM, CT, and AUA cases \footnote{This can be obtained by noting that in the AUA case, (\ref{zAUAD}) gives 
$z_{\rm AUA}^0(\tau) -z_{\rm AUA}^0(\tau') \approx 2 a^{-1} [n(\tau) - n(\tau')] \sinh (a\tau_p/4)$
if $|\tau -\tau'| \gg \tau_p$, when $n(\tau) - n(\tau')$ looks like $2\Delta/\tau_p$ in Fig. \ref{Avgf} if $\omega\tau_p/(2\pi) \ll 1$. }
while $f_{\bar{a}} \sim -[\omega^2/(\omega^2+ a_\circ^2)]\hbar /(2\pi \Delta)^2$ 
for the detector in circular motion ($\bar{a}=a_\circ $).
Furthermore all of them are negative in the domain with large $\Delta$. Therefore, for small $\Omega$ the effective temperatures in all of these cases will be significantly lower than the Unruh temperature of the UA detector in the high-acceleration regime. 

When the parameters $\bar{a}$, $\gamma$, $\Lambda_0$, $\Lambda_1$, $\Omega$, and $\omega$ are held fixed, we find the hierarchy of the effective temperatures CT $>$ SM $>$ AUA in the high-acceleration regime (see  Fig.~\ref{TC} (right)). 
This is consistent with what one can observe in Fig. \ref{Avgf}, where the value of $\int_0^\infty \bar{f}_{\bar{a}} d\Delta$ is CT $>$ SM $>$ AUA 
(all are negative).

\subsection{On-Resonance Cases}

\begin{figure}[h]
\includegraphics[width=4.8cm]{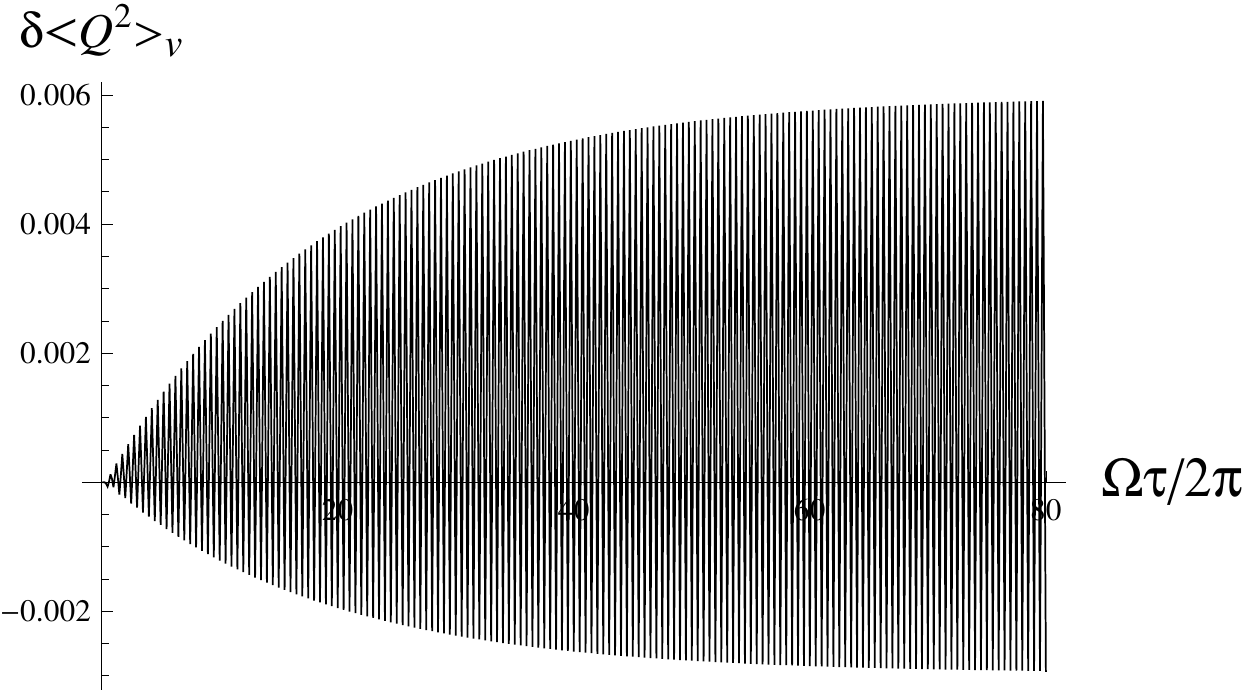} 
\includegraphics[width=4.8cm]{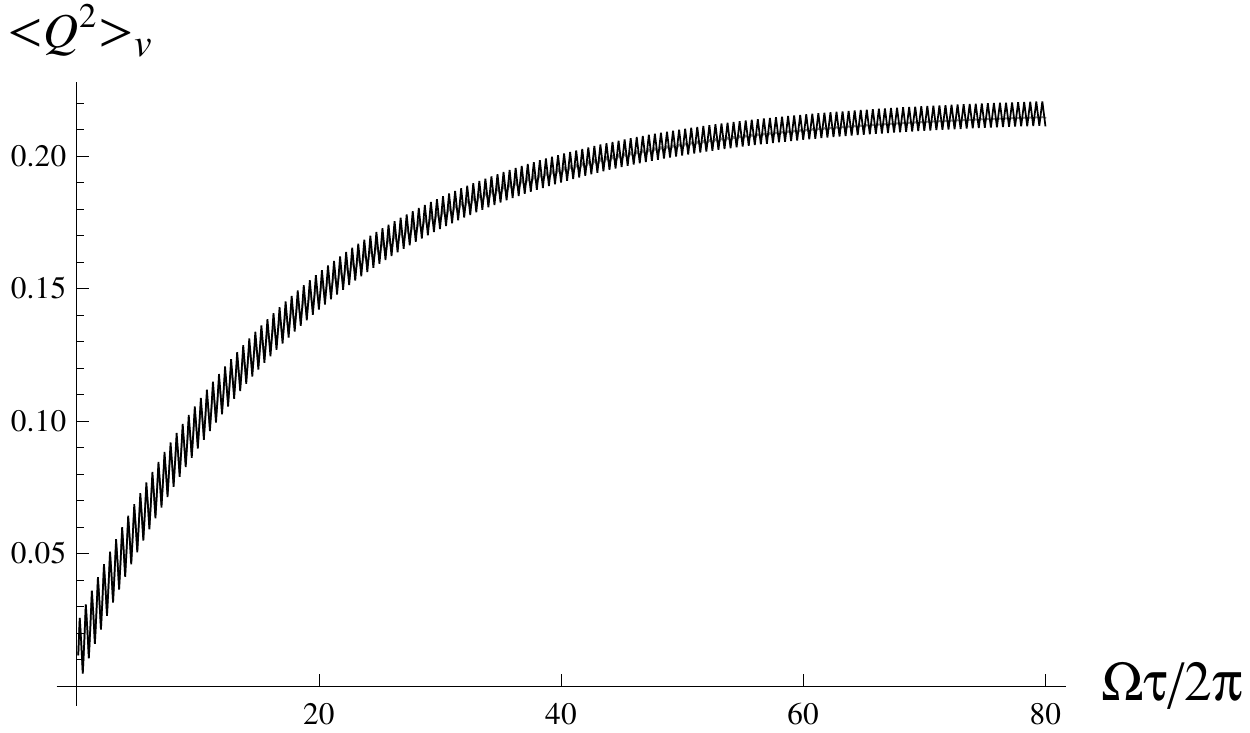}  
\includegraphics[width=4.8cm]{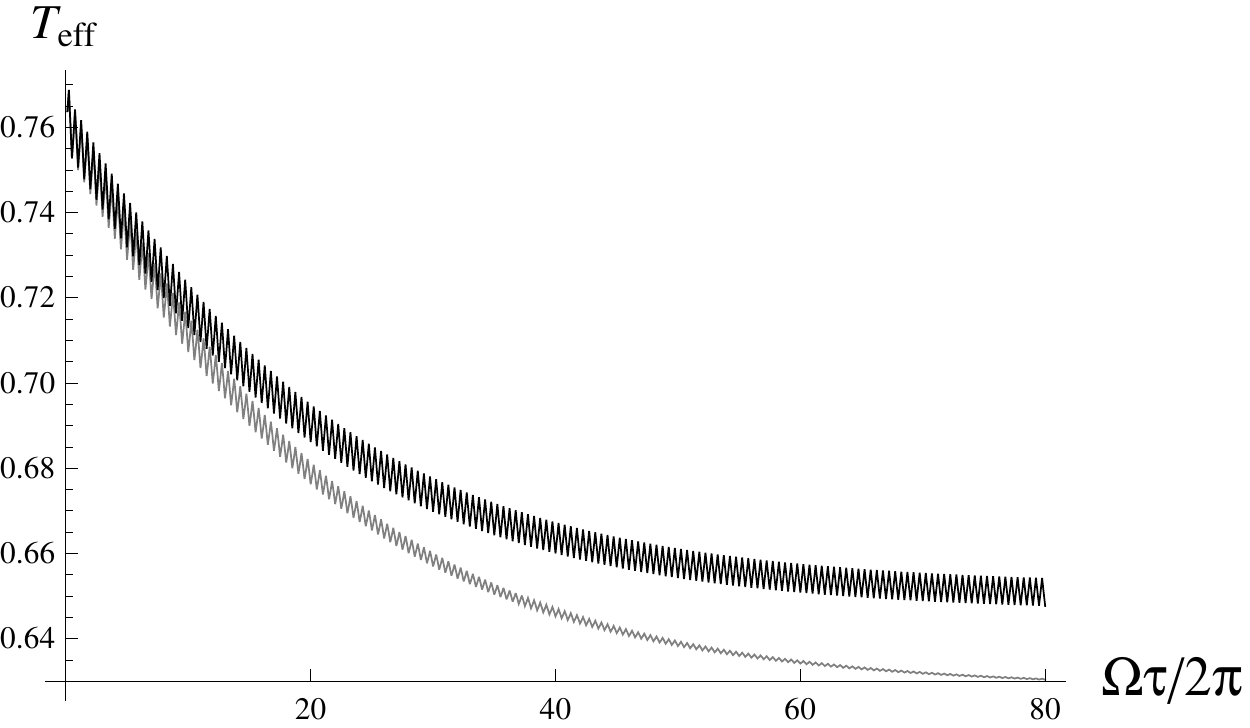}
\caption{Numerical results of an on-resonance case for the AUA detector (dark lines), compared with the results 
for the UA detector (grey lines) with the same $\bar{a}$. 
The values of the parameters are $\gamma=0.01$, $\Omega=2.3$, $a=2$, $\hbar=m_0=1$, and $\tau_0=0$.
Here the period of the alternating acceleration $\tau_p$ is the same as the natural period of the detector $2\pi/\Omega$.}
\label{Q2vResult2}
\end{figure}

One may wonder whether the corrections to the correlators  and the effective temperature may be amplified in the cases when
the detector's natural oscillation is on resonance with the oscillatory motion. Indeed, 
we have observed numerically that when $\Omega \approx 2\pi/\tau_p$, that is, when the natural frequency of the detector
is the same as the frequency of its oscillatory motion, the amplitudes of the oscillation of the subtracted correlators 
$\delta\langle {\cal R}_i, {\cal R}_j\rangle$ in time are maximized. 
In Fig. \ref{Q2vResult2} we give an example for the AUA detector with $\Omega = 2\pi/\tau_p$. One can see that
$\delta\langle \hat{Q}^2\rangle_{\rm v}$ 
oscillates significantly between positive and negative values while the mean values are positive.
However, the amplitudes are still small ($O(\gamma)$) compared with the total values of the v-part of the correlators
$\langle \hat{Q}^2\rangle_{\rm v}$. 
Note that in Fig. \ref{Q2vResult2}, the value of the proper acceleration $a=2$ is not large compared with other parameters. 
For $a \gg c\Omega$, the mean values of $\delta\langle \hat{Q}^2\rangle_{\rm v}$ and $\delta\langle \hat{P}^2\rangle_{\rm v}$ 
will become negative, too.

\subsection{Detecting the Unruh Effect in Oscillatory Motion?}

Since it is impossible to realize any eternal, ideal uniform acceleration in laboratory, many authors have suggested attempting to detect the Unruh effect on charges or atoms in linear oscillatory motion \cite{CT99, MFM11}. For example, Chen and Tajima have proposed 
detecting the Unruh effect on electrons from the radiation emitted by the electrons driven by a linearly polarized laser field \cite{CT99}.
It was implicitly assumed in these calculations that the autocorrelation of the UA worldline could be recovered from the autocorrelation of the CT or AUA worldline in the high acceleration limit. In the above results, however, we have shown that this is not the case, and furthermore that the steady-state average-temperature experienced by a CT or AUA detector differs from that of a UA detector. It then follows that the proposed signal of the Unruh effect emitted by the electrons or registered in the Berry phase of the atoms \cite{MFM11} under CT or AUA motion will also be modified by taking into account these differences.

In fact, we see quite generally that the steady state average temperature of any detector that travels along a trajectory whose spatial distance is bounded in time will behave more similarly to a circularly moving detector than a UA detector in the high acceleration limit if the time scale of the 
response of the detector is longer than the period of the oscillatory motion, i.e. $\Omega \lesssim \omega$ (note that in Fig.~\ref{TC} (left), the effective temperature in the AUA case becomes lower than the Unruh temperature at high acceleration limit even when $\Omega=2.3 > 1 = \omega$, and the one for the CT has the same tendency). From a purist's point of view, this makes verification of the Unruh effect considerably more challenging. However, from a pragmatist's point of view, since the mechanisms by which the vacuum fluctuations excite both oscillating detectors and UA detectors are the same, experimental verification of the temperature dependence of the kind we have found would surely add weight to the evidence in favour of the Unruh effect.

\section{Conclusion}
\label{sec:conclusion}

We have explored the non-equilibrium effects of vacuum noise under oscillatory motion using a formally exactly solvable model of a harmonic oscillator detector. This model provides an important tool that can be used to analyze the spectra of vacuum fluctuations even in non-stationary situations. Since the model allows us to analyze the time-evolution of the detector exactly, we were also able to investigate the approach to equilibrium that a UA detector makes once it is switched on at some finite time finding that it obeys Newton's cooling equation at near-equilibrium temperatures.

We have emphasized that the instantaneous proper acceleration does not on its own determine the effective temperature experienced by a detector at late times. Rather, the late time temperature behaviour is more strongly dependent on the geometry of the worldline. More specifically, since oscillating detectors like circularly moving detectors have spatial trajectories that are bounded over time inside a finite spatial region, their steady state temperature behaviour is more similar to each other than to that of the Unruh temperature if the time scale of the detector's response is longer than the period of the oscillatory motion.  

We have taken several steps to remove detector dependent features which have the potential to significantly change the state field that the detector is attempting to measure. In particular, we worked in a regime where the UA detector recovered the Unruh temperature at late times -- the large acceleration limit. Our results indicate that at large oscillation-frequencies and large average-accelerations the temperature of an oscillating detector is lower than that of the Unruh temperature. 

The exact behavior of any accelerated detector will depend on the features of the detector model one uses. However, since the Wightman functions will be the same (for the same choice of field), we expect that the model and numerical analysis we have presented here will provide a good description for most experimental proposals that involve oscillatory motion. 

Our work is also commensurate with recent work that considers the physics of accelerating detectors in cavities with various boundary conditions \cite{C1,C2}. In this case the Unruh effect is again confirmed, provided a sufficient number of modes in the cavity are taken into account; furthermore, thermalization can non-perturbatively be demonstrated to hold  \cite{C3}.

\begin{acknowledgments}

JD acknowledges financial support from EPSRC under the CAF Grant EP/G00496X/2 to I. Fuentes.
SYL is supported by the NSC Taiwan under the grants 99-2112-M-018-001-MY3, 102-2112-M-018-005-MY3, 
and in part by the National Center for Theoretical Sciences, Taiwan. He acknowledges the hospitality of
the Department of Physics, University of Queensland during the RQI5 workshop and
the Department of Physics and Astronomy, University of Waterloo during the RQI-N 2012 workshop.
BLH was supported by NSF Grant PHY-0801368 when this work commenced. 
He acknowledges fruitful collaborations with Alpan Raval and Don Koks which established the nonequilibrium approach 
to particle motion in quantum fields and lively discussions with Phil Johnson at the 2000 Capri meeting which resulted 
in \cite{HJ00} and related reports for that meeting. 
RBM was supported in part by the Natural Sciences and Engineering Research Council of Canada.
\end{acknowledgments}

\appendix

\section{Late-time effective temperature of detector in circular motion}
\label{sec:circ}

By virtue of the stationary property of both (\ref{WFcirc}) and (\ref{WFUAD}), obtaining the late-time results for the subtracted correlators can be simplified to an one-dimensional integral:
\begin{eqnarray}
  & &\lim_{\gamma\tau\to\infty}\delta \left<\right. \hat{Q}^2(\tau)\left.\right>_{\rm v} = \lim_{\gamma\tau\to \infty}
  {\lambda_0^2\over m_0^2 \Omega^2} {\rm Re}\int_{0}^\tau d\tilde{\tau} \int_{0}^{\tau} d\tilde{\tau}'
   K(\tau - \tilde{\tau}) K(\tau - \tilde{\tau}')f_{a_\circ}(\tilde{\tau}-\tilde{\tau}') \nonumber\\
  &=& \lim_{\gamma\tau\to \infty}
  {8\pi\gamma\over m_0\Omega^2} {\rm Re} \int_{-\tau}^\tau d\Delta \int_{|\Delta|\over 2}^{\tau-{|\Delta|\over 2}} d{\cal T}
   {e^{-2\gamma(\tau-{\cal T})}\over 2}\left[\cos \Omega\Delta - \cos 2\Omega(\tau-{\cal T})\right] f_{a_\circ}(\Delta) \nonumber\\
  &=& {4\pi\over m_0\Omega_r^2} \lim_{\gamma\tau\to \infty} 
  \int_{0}^\tau d\Delta  f_{a_\circ}(\Delta)\left\{
  e^{-\gamma\Delta}\left(\cos\Omega\Delta+{\gamma\over \Omega}\sin\Omega\Delta\right)+O(e^{-\gamma\tau})\right\},
\label{dQ2vstation}
\end{eqnarray}
where the $O(e^{-\gamma\tau})$ terms can be neglected at late times.
Similar arguments can be applied to $\delta \left<\right. \hat{P}^2(\tau)\left.\right>_{\rm v}$ to get
\begin{equation}
  \lim_{\gamma\tau\to\infty}\delta \left<\right. \hat{P}^2(\tau)\left.\right>_{\rm v} 
  = 4\pi m_0 \lim_{\gamma\tau\to \infty} \int_{0}^\tau d\Delta  f_{a_\circ}(\Delta)\left\{
  e^{-\gamma\Delta}\left(\cos\Omega\Delta-{\gamma\over \Omega}\sin\Omega\Delta\right)+O(e^{-\gamma\tau})\right\},
\end{equation}
and $\delta \left<\right. \hat{Q}(\tau),\hat{P}(\tau)\left.\right>_{\rm v}$ vanishes at late times.
From these subtracted correlators one can obtain the late-time effective temperatures, which depends explicitly
on the natural frequency of the detector $\Omega$.

\begin{figure}[h]
\includegraphics[width=7cm]{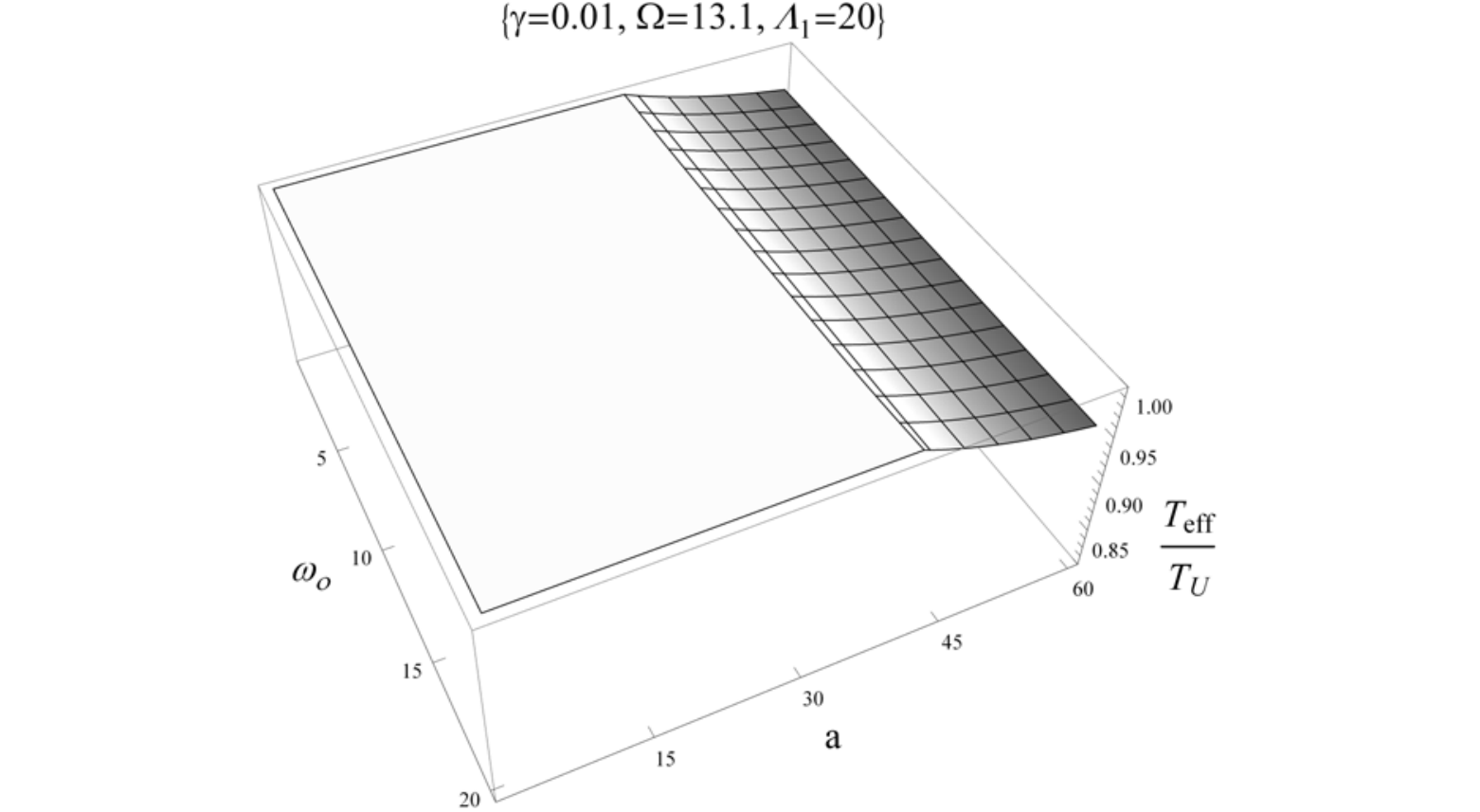} 
\includegraphics[width=7cm]{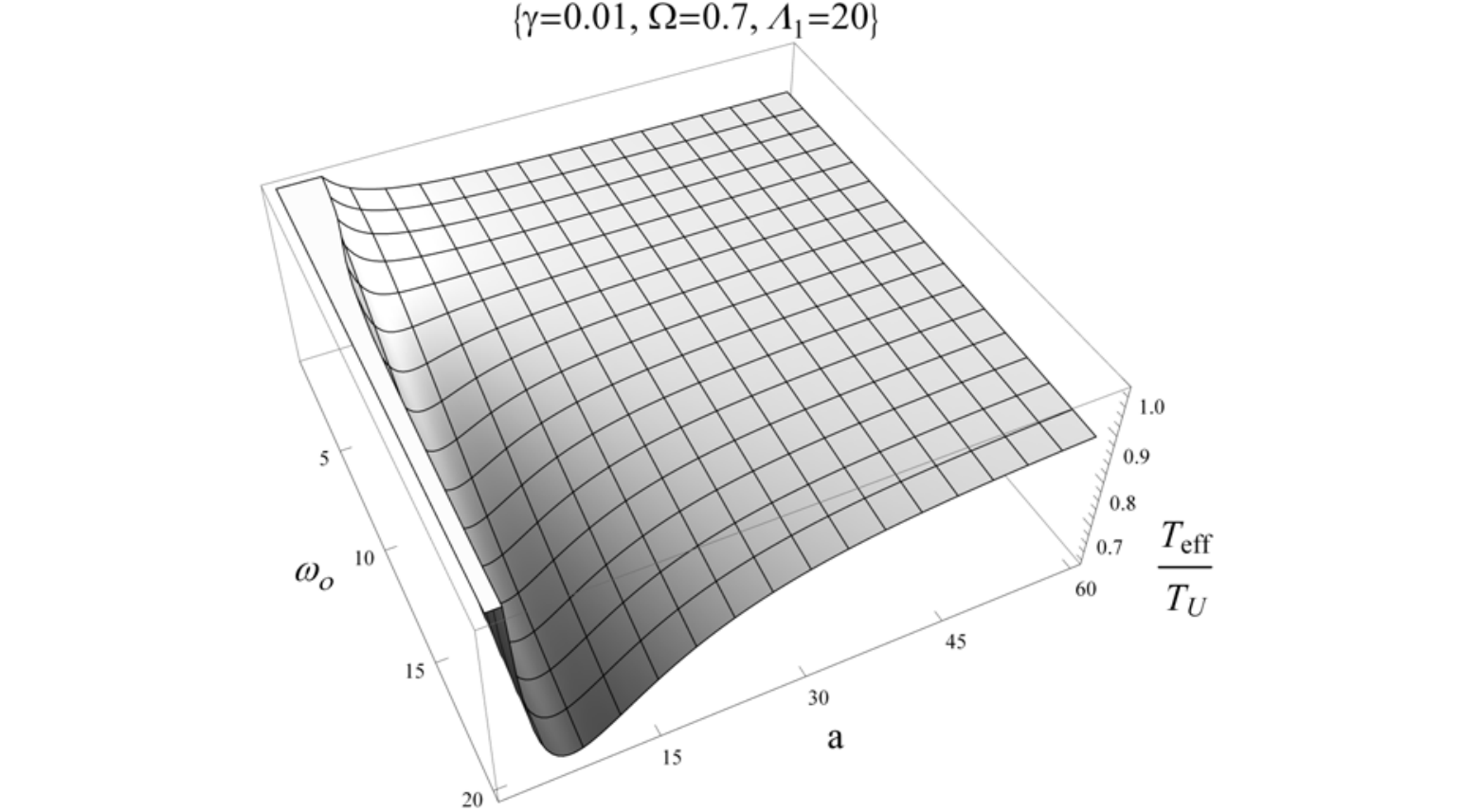}  
\center{\includegraphics[width=7cm]{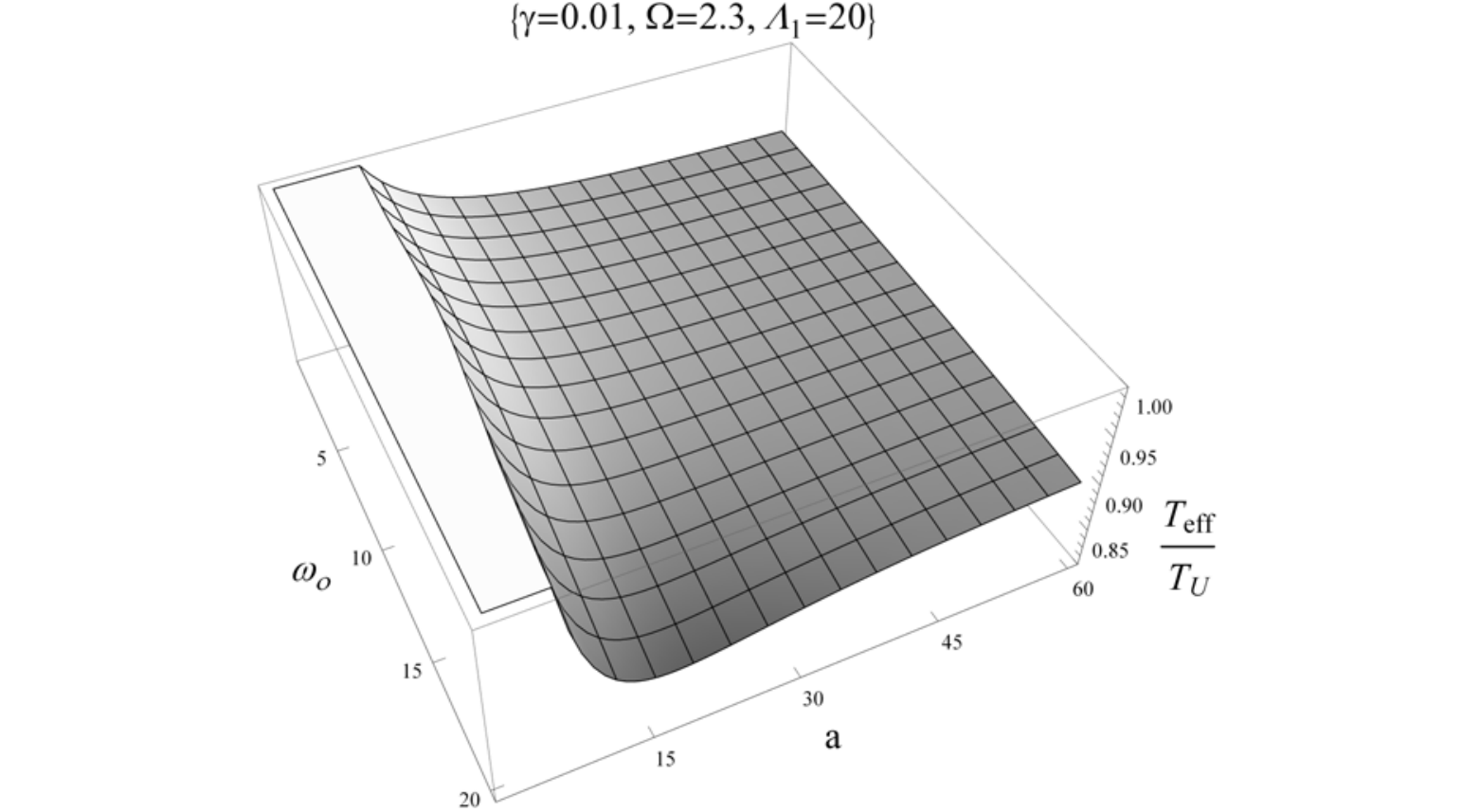}}   
\caption{The ratio of the effective temperature $T_{\rm eff}$ for a detector
in circular motion to the Unruh temperature $T_U$ of a uniformly, linearly accelerated detector with the same proper acceleration
$a_\circ=\Gamma v \omega$ as a function of $\omega$ and $a_\circ$. Only the values less than $1$ are shown
to manifest the border line with $T_{\rm eff}=T_U$.}
\label{Teff2TU}
\end{figure}

Similar to the detectors in oscillatory motion, the effective temperature experienced by a detector in circular motion
can also be greater or less than the effective temperature in a UA detector with the same proper acceleration.
In Fig. \ref{Teff2TU} we compare the late-time effective temperature of the circularly moving detector with that of the UA detector 
at the same acceleration. In the high acceleration regime ($a_\circ \gg \Omega$, $\omega$) the effective temperature $T_{\rm eff}$ 
roughly depends on the proper acceleration $a_{\circ}$ only. 
For extremely large accelerations in our results, $T_{\rm eff}\approx 0.91 T_U = 0.91 a_\circ/2\pi$,
independent of the values of $\Omega$ and $\omega$.
This is consistent with the numerical results of Bell and Leinaas in Fig. 2 of \cite{BL83} for $|g| \to 0$
\footnote{The energy difference in the detector $\hbar\Omega$, which corresponds to $\Delta$ in \cite{BL83} is less than $\hbar a_\circ/ c\Gamma$ in our high acceleration limit.}, and the analytic result $T_{\rm eff} = a_\circ/4\sqrt{3}$ given by Unruh in \cite{Unr98}.

In Fig. \ref{Teff2TU} (right), one can also see that in the regime with large $\omega$, small $\Omega$, and not-too-small $a_\circ$,
the effective temperature $T_{\rm eff}$ can be much lower than $T_U$.

\begin{thebibliography}{99}

\bibitem{Unr76}  W. G. Unruh, 
{\it Notes on black hole evaporation},
{\it Phys. Rev.} {\bf D 14} (1976) 870.

\bibitem{DeW79} B. S. DeWitt, 
{\it Quantum gravity: the new synthesis}, 
in {\it General Relativity: an Einstein Centenary Survey}, 
edited by S. W. Hawking and W. Israel, Cambridge University Press, Cambridge (1979).

\bibitem{Haw75} S. W. Hawking, 
{\it Particle creation by black holes}, 
{\it Commun. Math. Phys.} {\bf 43} (1975) 199.

\bibitem{RHA}  A. Raval, B. L. Hu and J. Anglin,
{\it Stochastic theory of accelerated detectors in a quantum field},
{\it Phys. Rev.} {\bf D 53} (1996) 7003.

\bibitem{RHK}  A. Raval, B. L. Hu and D. Koks, 
{\it Near-thermal radiation in detectors, mirrors, and black holes: A stochastic approach},
{\it Phys. Rev.} {\bf D 55} (1997) 4795.

\bibitem{JH1}  P.~R. Johnson and B.~L. Hu, 
{\it Stochastic theory of relativistic particles moving in a quantum field:  
Scalar Abraham-Lorentz-Dirac-Langevin equation,	radiation reaction and vacuum fluctuations},   
{\it Phys. Rev.} {\bf D 65} (2002) 065015. 

\bibitem{JHFoP}  P.~R. Johnson and B.~L. Hu, 
{\it Unruh Effect in a uniformly accelerated charge:  From quantum fluctuations to classical radiation},  
{\it Found. Phys.} {\bf 35} (2005) 1117. 

\bibitem{ifctp}  
R. Feynman and F. L. Vernon Jr., 
{\it The theory of a general quantum system interacting with a linear dissipative system},
{\it Annals Phys.} {\bf 24} (1963) 118; 
J. Schwinger, 
{\it Brownian motion of a quantum oscillator}, 
{\it J. Math. Phys.} {\bf 2} (1961) 407; 
L. V. Keldysh, 
{\it Diagram technique for nonequilibrium processes},
{\it Zh. Eksp. Teor. Fiz.} {\bf 47} (1964) 1515 [Engl. trans. {\it Sov. Phys. JEPT} {\bf 20} (1965) 1018].

\bibitem{cgea}  B. L. Hu, 
{\it Coarse graining and back reaction in inflationary and minisuperspace cosmology},
in {\it Relativity and Gravitation: Classical and Quantum}, 
Proc. SILARG VII, Cocoyoc, Mexico 1990. eds. J. C. D' Olivo et al.,World Scientific, Singapore (1991). 
Reproduced in 
E. A. Calzetta, B. L. Hu, and F. D. Mazzitelli,	
{\it Coarse-grained effective action and renormalization group theory in semiclassical gravity and cosmology}, 	
{\it Phys. Rep.} {\bf 352} (2001) 459. 

\bibitem{CalHu08}
E. Calzetta  and B. L. Hu, 
{\it Nonequilibrium Quantum Field Theory}, Cambridge University Press, Cambridge, (2008).

\bibitem{HJ00} B. L. Hu and P. R. Johnson, 
{\it Beyond Unruh effect: nonequilibrium quantum dynamics of moving charges}, 
in \cite{Capri00}. 

\bibitem{Capri00}
{\it Proceedings of the Capri Workshop on Quantum Aspect of Beam Physics, Oct. 2000}, edited by Pisin Chen, 
World Scientific, Singapore (2001).

\bibitem{OtherExpProp}
J. Rogers,
{\it Detector for the temperaturelike effect of acceleration},
{\it Phys. Rev. Lett.} {\bf 61} (1988) 2113; 
M. O. Scully, V. V. Kocharovsky, A. Belyanin, E. Fry, and F. Capasso, 
{\it Enhancing Acceleration Radiation from Ground-State Atoms via Cavity Quantum Electrodynamics}, 
{\it Phys. Rev. Lett.} {\bf 91} (2003) 243004; {\bf 93} (2004) 129302;
B. L. Hu and A. Roura, 
{\it Phys. Rev. Lett.} {\bf 93} (2004) 129301; 
R. Sch\"utzhold, G. Schaller, and D. Habs, 
{\it Signatures of the Unruh effect from electrons accelerated by ultrastrong laser fields},
{\it Phys. Rev. Lett.} {\bf 97} (2006) 121302;
G. E. A. Matsas and D. A. T. Vanzella, 
{\it Decay of protons and neutrons induced by acceleration},
{\it Phys. Rev.} {\bf D 59} (1999) 094004. 

\bibitem{CT99} P. Chen and T. Tajima, 
{\it Testing Unruh radiation with ultraintense lasers},
{\it Phys. Rev. Lett.} {\bf 83} (1999) 256. 

\bibitem{MFM11} E. Martin-Martinez, I. Fuentes, and R. B. Mann, 
{\it Using Berrys's phase to detect the Unruh effect at lower accelerations},
{\it Phys. Rev. Lett.} {\bf 107} (2011) 131301.

\bibitem{BL83} J. S. Bell and J. M. Leinaas, 
{\it Electrons as accelerated thermometers},
{\it Nucl. Phys.} {\bf B 212} (1983) 131.

\bibitem{BL87}J. S. Bell and J. M. Leinaas, 
{\it The Unruh effect and quantum fluctuations of electrons in storage rings},
{\it Nucl. Phys.} {\bf B 284} (1987) 488.

\bibitem{Len00}  J. M. Lennias, 
{\it Unruh effect in storage rings}, 
in \cite{Capri00}.

\bibitem{Unr98}  W. G. Unruh,
{\it Acceleration radiation for orbiting electrons}, in 
{\it Monterey Workshop on Quantum Aspects of Beam Physics}, edited by Pisin Chen
World Scientific, Singapore (1998); 
{\it Phys. Rep.} {\bf 307} (1998) 163. 

\bibitem{Doukas}
J. Doukas and B. Carson, 
{\it Entanglement of two qubits in a relativistic orbit},
{\it Phys. Rev.} {\bf A 81} (2010) 062320;
J. Doukas and L. C. L. Hollenberg,
{\it Loss of spin entanglement for accelerated electrons in electric and magnetic fields},
{\it Phys. Rev.} {\bf A 79} (2009) 052109.

\bibitem{SS92} B. F. Svaiter and N. F. Svaiter, 
{\it Inertial and noninertial particle detectors and vacuum fluctuations},
{\it Phys. Rev.} {\bf D 46} (1992) 5267; {\bf D 47} (1993) 4802.

\bibitem{SP96} L. Sriramkumar and T. Padmanabhan,
{\it Response of finite time particle detectors in noninertial frames and curved space-time},
{\it Class. Quantum Grav.} (1996) {\bf 13}, 2061.

\bibitem{Sc04} A. Schlicht,
{\it Considerations on the Unruh effect: Causality and regularization},
{\it Class. Quantum Grav.} {\bf 21} (2004) 4647.

\bibitem{LS06} J. Louko and A. Satz,
{\it How often does the Unruh-DeWitt detector click? Regularisation by a spatial profile},
{\it Class. Quantum Grav.} {\bf 23} (2006) 6321.

\bibitem{Sa07} A. Satz,
{\it Then again, how often does the Unruh-DeWitt detector click if we switch it carefully?}
{\it Class. Quantum Grav.} {\bf 24} (2007) 1719.

\bibitem{OM07} N. Obadia and M. Milgrom,
{\it On the Unruh effect for general trajectories},
{\it Phys. Rev.} {\bf D 75} (2007) 065006.

\bibitem{KP10} D. Kothawala and T. Padmanabhan,
{\it Response of Unruh-DeWitt detector with time-dependent acceleration},
{\it Phys. Lett.} {\bf B 690} (2010) 201.

\bibitem{BV12} Barbado and Visser, 
{\it Unruh-DeWitt detector event rate for trajectories with time-dependent acceleration},
{\it Phys. Rev.} {\bf D 86} (2012) 084011.

\bibitem{LH06}
S.-Y. Lin and B. L. Hu, 
{\it Accelerated detector-quantum field correlations: From vacuum fluctuations to radiation flux},
{\it Phys. Rev.} {\bf D 73} (2006) 124018.

\bibitem{LH07}
S.-Y. Lin and B. L. Hu, 
{\it Backreaction and the Unruh effect: New insights from exact solutions of uniformly accelerated detectors},
{\it Phys. Rev.} {\bf D 76} (2007) 064008.

\bibitem{OLMH12} D. C. M. Ostapchuk, S.-Y. Lin, R. B. Mann, and B. L. Hu,
{\it Entanglement dynamics between inertial and non-uniformly accelerated detectors}, 
{\it JHEP} {\bf 07} (2012) 072. 

\bibitem{TAKAGI}
S. Takagi, 
{\it Vacuum noise and stress induced by uniform acceleration},
{\it Prog. Theo. Phys. Supp.} {\bf 88} (1986) 1.

\bibitem{Sciama81}
D. W. Sciama, P. Candelas, and D. Deutsch, 
{\it Quantum field theory, horizons and thermodynamics},
{\it Adv. Phys.} {\bf 30} (1981) 327.

\bibitem{BD82} N. D. Birrell and P. C. W. Davies, 
{\it Quantum Fields in Curved Space}, 
Cambridge University Press, Cambridge (1982).

\bibitem{Berry}
E. Martin-Martinez, A. Dragan, R. B. Mann, I. Fuentes, 
{\it Berry phase quantum thermometer},
{\it New J. Phys.} {\bf 15} (2013) 053036.

\bibitem{C1} D. E. Bruschi, A. R. Lee, and I. Fuentes, 
{\it Time evolution techniques for detectors in relativistic quantum information},
{\it J. Phys.} {\bf A 46} (2013) 165303. 

\bibitem{C2} E. G. Brown, E. Martin-Martinez, N. C. Menicucci, and R. B. Mann,
{\it Detectors for probing relativistic quantum physics beyond perturbation theory},
{\it Phys. Rev.} {\bf D 87} (2013) 084062.

\bibitem{C3} W.G. Brenna, E.G. Brown, R.B. Mann and E. Martin-Martinez, 
{\it Universality and thermalization in the Unruh Effect},
{\it Phys. Rev.} {\bf D 88} (2013) 064031.

\end{thebibliography}
\end{document}